\documentclass{article} 
\usepackage{iclr2027_conference,times}

\usepackage[table]{xcolor}
\usepackage{hyperref}
\usepackage{url}
\usepackage{xspace}
\usepackage{amsmath,amsfonts,amssymb,bm}
\usepackage{booktabs}
\usepackage{multirow}
\usepackage{enumitem}
\usepackage{pifont}
\usepackage{graphicx}
\usepackage{wrapfig}
\usepackage{float}
\usepackage{placeins}

\newcommand{\methodname}{OPTD\xspace}
\newcommand{\yesmark}{\textcolor{green!50!black}{\ding{51}}}
\newcommand{\nomark}{\textcolor{red!70!black}{\ding{55}}}
\newcommand{\levelbullet}{\textcolor{blue!55!black}{\raisebox{0.1ex}{\scriptsize\ding{108}}}\hspace{0.45em}}

\def\Figref#1{Figure~\ref{#1}}

\def\Secref#1{Section~\ref{#1}}

\def\eqref#1{equation~\ref{#1}}

\def\1{\bm{1}}
\newcommand{\train}{\mathcal{D}}

\def\eps{{\epsilon}}

\DeclareMathAlphabet{\mathsfit}{\encodingdefault}{\sfdefault}{m}{sl}
\SetMathAlphabet{\mathsfit}{bold}{\encodingdefault}{\sfdefault}{bx}{n}

\newcommand{\E}{\mathbb{E}}
\newcommand{\Ls}{\mathcal{L}}

\newcommand{\softmax}{\mathrm{softmax}}

\newcommand{\KL}{D_{\mathrm{KL}}}

\author{%
\normalfont
\begin{tabular*}{0.96\textwidth}[t]{@{\extracolsep{\fill}}lllll@{}}
\textbf{Jujia Zhao}\textsuperscript{1,*} &
\textbf{Jiakai Tang}\textsuperscript{2,*} &
\textbf{Jing Yu}\textsuperscript{3,*} &
\textbf{Wen Chen}\textsuperscript{3,\(\dagger\)} &
\textbf{Jian Wu}\textsuperscript{3} \\
\textbf{Han Zhu}\textsuperscript{3} &
\textbf{Fei Sun}\textsuperscript{4} &
\textbf{Suzan Verberne}\textsuperscript{1} &
\textbf{Zhaochun Ren}\textsuperscript{1,\(\dagger\)} &
\\[0.55em]
\multicolumn{5}{@{}l@{}}{%
\textsuperscript{1}Leiden University
\quad
\textsuperscript{2}Renmin University of China
\quad
\textsuperscript{3}Alibaba Group} \\
\multicolumn{5}{@{}l@{}}{%
\textsuperscript{4}Institute of Computing Technology, Chinese Academy of Sciences}
\end{tabular*}
}

\iclrfinalcopy

\title{UserArena: Benchmarking Interactive User Simulation for Long-Horizon Shopping Agents}

\begin{document}

\maketitle
\begingroup
\renewcommand{\thefootnote}{\fnsymbol{footnote}}
\footnotetext[1]{Equal contribution. Primary contact:
\href{mailto:zhao.jujia.0913@gmail.com}{zhao.jujia.0913@gmail.com}.}
\footnotetext[2]{Corresponding authors.}
\endgroup
\lhead{}

\begin{abstract}
Agentic recommendation pursues shopper goals through sequences of context-dependent decisions.
Developing and evaluating such policies requires faithful recommendation world models that capture both the interactive shopping environment and users' behavioral responses over long interactions.
Existing resources rarely provide long-horizon, fine-grained real-user trajectories or support trajectory-level evaluation and offline validation against production A/B results.
We introduce \textbf{\textsc{UserArena}}, a real-world e-commerce interactive benchmark containing 64,729 session trajectories from 11,010 users, spanning 566,102 items and 13 action types.
\textsc{UserArena} pairs these trajectories with an interactive environment and supports offline validation against outcomes from a controlled production A/B test, enabling five-level evaluation from output validity and local decisions to sequential behavior, personalization, and outcomes.
To improve simulator fidelity over these long-horizon interactions, we further propose \textbf{O}n-\textbf{P}olicy \textbf{T}rajectory \textbf{D}istillation (\textbf{\methodname}), which uses trajectory-level intent to provide dense teacher targets on student-generated rollouts.
Experiments show that \methodname achieves the strongest overall fidelity among the evaluated simulators under \textsc{UserArena}'s evaluation protocol and most closely reproduces the production policy effect in offline simulation.\footnote{Our code and data are available at \url{https://github.com/RecGPT-Team/UserArena}. The project page is available at \url{https://recgpt-team.github.io/UserArena-Page}.}
\end{abstract}

\section{Introduction}
\label{sec:introduction}

Recent advances in language and multimodal agents create new opportunities for agentic recommendation, in which an agent pursues a shopper's goal through a sequence of context-dependent decisions~\citep{recoworld,agentrecbench}.
Post-training and evaluation of the underlying recommendation policy require a scalable \emph{recommendation world model} that simulates how the shopping environment and its users respond to each policy action~\citep{recsim}.
An executable environment models what the policy can observe and change, while a user simulator models how shoppers respond to those changes~\citep{webshop,zhou2024webarena,recagent,agent4rec}.
Together, they support offline and counterfactual evaluation as well as reinforcement-learning-based optimization, reducing reliance on costly online experiments~\citep{virtualtaobao,kuaisim}.

\begin{figure}[t]
\centering
\includegraphics[width=\textwidth]{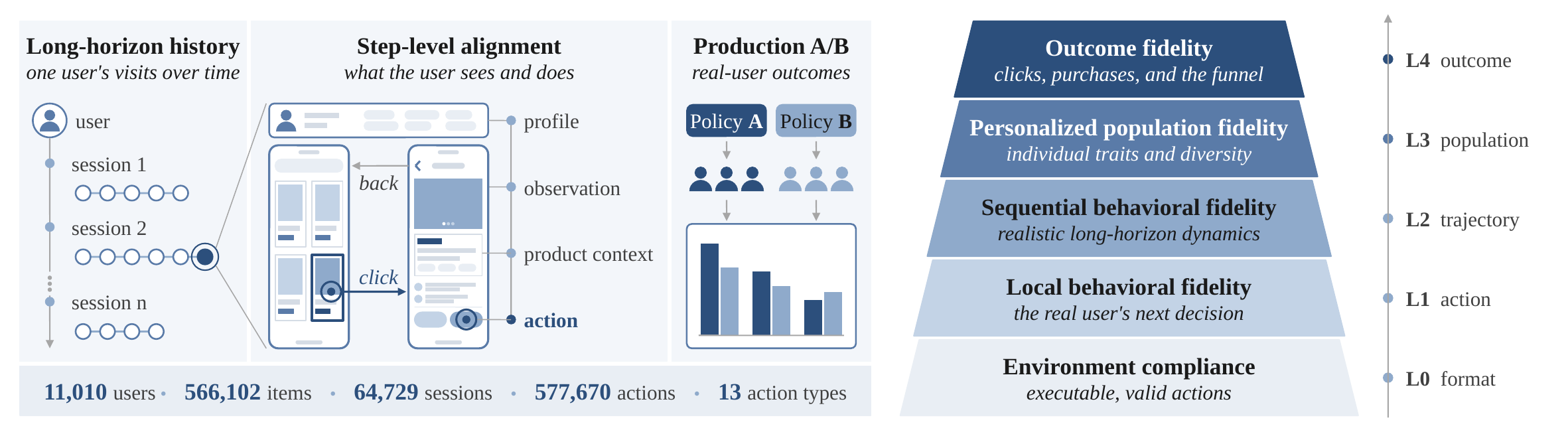}
\vspace{-14pt}
\caption{Overview of \textsc{UserArena}. \textbf{Left:} cross-session histories, step-aligned interaction records, production A/B validation, and dataset scale. \textbf{Right:} the five-level evaluation protocol, progressing from environment compliance (L0) to outcome fidelity (L4).}
\label{fig:eval_levels}
\end{figure}

Existing resources provide rich behavioral data and interactive evaluation for such world models.
Amazon Reviews~\citep{Hou2026Bridging} contains ratings and review text, Qilin~\citep{Chen2025Qilin} adds multimodal content and app-level sessions, and OmniBehavior~\citep{omnibehavior} captures longitudinal traces.
RecBench+~\citep{Huang2026RecBenchPlus} and OPeRA~\citep{opera} move from offline records toward interactive evaluation.
Yet two crucial gaps remain: 
\textbf{At the data level,} existing datasets rarely provide long-horizon trajectories that align real-user profiles, visual observations, product context, and actions across repeated sessions.
\textbf{At the evaluation level,} next-action agreement measures step-wise choices in isolation but does not assess whether simulated trajectories preserve a user's behavioral tendencies over long horizons.
Existing benchmarks also rarely test whether simulator-based policy comparisons agree with production A/B outcomes.
These gaps limit assessment of long-horizon behavioral fidelity and the real-world validity of simulation conclusions.

To address these limitations, we introduce \textbf{\textsc{UserArena}}, an interactive benchmark combining a production-grounded shopping environment with real-user trajectories recorded on one of Asia's largest e-commerce platforms (\Figref{fig:eval_levels}).
It contains real-user data comprising \textbf{11,010 users, 566,102 items, and 64,729 session trajectories} and covers a complete action space of 13 action types.
Each trajectory aligns user profiles, visual observations, product context, and actions at every step.
For long-horizon evaluation, \textsc{UserArena} links sessions from the same user into cross-session trajectories.
It also supports offline validation by comparing the same recommendation-policy pair in simulation with outcomes from a controlled production A/B test.
We evaluate simulators at five levels: output validity (\textbf{L0}), local behavioral fidelity (\textbf{L1}), sequential behavioral fidelity (\textbf{L2}), personalization fidelity (\textbf{L3}), and outcome fidelity (\textbf{L4}).
The protocol traces fidelity from output validity and local decisions to long-horizon behavior, personalization, and aggregate outcomes.

Beyond constructing the benchmark, we also address the challenge of learning faithful user simulators.
Simulator design has progressed from hand-crafted and log-driven models to LLM-based models~\citep{recsim,virtualtaobao,recagent,agent4rec}, yet reproducing user behavior over long rollouts remains difficult for three reasons.
\textbf{(i) Behavioral miscalibration:} learned simulators can deviate from real user behavior~\citep{agent4rec,yoon2024evaluating}.
\textbf{(ii) Sparse supervision:} exact next-action targets penalize plausible alternatives and overlook trajectory-level quality~\citep{wang2025customerr1,zhang2026shopr1}.
\textbf{(iii) Exposure bias:} teacher forcing conditions only on ground-truth prefixes, creating a mismatch with autoregressive inference~\citep{dai2021aicm,chi2022offline}.
To address these challenges, we propose \textbf{O}n-\textbf{P}olicy \textbf{T}rajectory \textbf{D}istillation (\textbf{\methodname}), which distills teacher guidance on current-policy responses at real states~\citep{agarwal2024onpolicy,zhao2026self}.
This on-policy supervision trains on student-generated prefixes, while trajectory-level intent guides teacher targets toward coherent long-horizon behavior.

In summary, our contributions are threefold.
(1) We introduce \textsc{UserArena}, a real-world e-commerce interactive benchmark pairing long-horizon user trajectories with an interactive environment and a five-level protocol spanning environment compliance, local decisions, sequential behavior, personalization, and outcomes.
(2) We propose \methodname, which uses trajectory-level intent and dense teacher targets to guide student-generated rollouts toward coherent long-horizon behavior.
(3) We benchmark representative text- and GUI-based simulators for effectiveness and efficiency and characterize \textsc{UserArena}'s behavioral composition and interaction horizons.

\section{The \textsc{UserArena} Benchmark}
\label{sec:benchmark}

\subsection{Data and Interaction Design}
\label{subsec:benchmark_design}

A credible e-commerce interactive benchmark requires a production-grounded environment that preserves what shoppers can observe and do.
Beyond this environmental foundation, two requirements are essential.
First, the benchmark should use fine-grained real-user trajectories that align user profiles, visual observations, product context, and actions while capturing behavior over long horizons.
Second, its evaluation should go beyond next-action agreement to assess complete trajectories and whether simulator-based policy comparisons agree with real-user outcomes in production.

Existing resources cover these requirements partially. 
E-commerce datasets contain real preferences and interaction logs but rarely retain the visual context of each decision~\citep{Harper2015MovieLens,Hou2026Bridging,Chen2025Qilin,rl4rs}. 
Interactive platforms enable closed-loop evaluation but depend on user dynamics whose fidelity remains insufficiently established~\citep{recsim,virtualtaobao,kuaisim}. 
Recent benchmarks improve behavioral grounding through visual shopping traces or longitudinal records~\citep{opera,omnibehavior,tang2026towards,zhang2026abagent}, yet none combines real step-level trajectories, visual context, a full shopping action space, long-horizon cross-session behavior, and production A/B validation (Table~\ref{tab:benchmark_comparison}).

\begin{table}[t]
\caption{Comparison of representative interactive recommendation benchmarks. Full Action Space covers the complete executable interaction loop. Long-Horizon denotes temporally ordered user histories spanning multiple sessions. Production A/B Validation evaluates recommendation policies offline in simulation and compares their effects with outcomes observed in production A/B tests.}
\label{tab:benchmark_comparison}
\centering
\footnotesize
\setlength{\tabcolsep}{4pt}
\resizebox{0.98\linewidth}{!}{%
\begin{tabular}{@{}lccccc@{}}
\toprule
Benchmark & Real Traces & Visual Context & Full Action Space & Long-Horizon & Production A/B \\
\midrule
Qilin & \yesmark & \nomark & \nomark & \nomark & \nomark \\
RL4RS & \yesmark & \nomark & \nomark & \nomark & \nomark \\
OPeRA & \yesmark & \yesmark & \yesmark & \nomark & \nomark \\
OmniBehavior & \yesmark & \nomark & \nomark & \yesmark & \nomark \\
USB & \yesmark & \yesmark & \yesmark & \nomark & \nomark \\
A/B Agent & \nomark & \yesmark & \yesmark & \nomark & \nomark \\
\textbf{\textsc{UserArena}} & \yesmark & \yesmark & \yesmark & \yesmark & \yesmark \\
\bottomrule
\end{tabular}
}
\vspace{-1em}
\end{table}

To bridge this gap, \textsc{UserArena} combines a production-grounded shopping environment, fine-grained real-user trajectories, and an evaluation protocol spanning trajectory fidelity and production validation.
The environment reconstructs production recommendation-feed and item-detail pages with 13 action types, from feed navigation and product inspection to add-to-cart and purchase.
It updates the visible feed by cropping the corresponding viewport from the pre-rendered full feed as the simulator scrolls.
Clicking any currently exposed item, including one not clicked by the real user, opens an interactive detail page reconstructed from that item's pre-rendered views and recorded product content.
Following the observation--action formulation of interactive agents~\citep{webshop,zhou2024webarena,opera,zhang2025sta}, each recorded decision aligns the shopper's profile, visual observation, visible product context, and action.
The benchmark contains \textbf{11,010 users, 64,729 sessions, 577,670 actions, and 566,102 products} and 6 user attributes.
Across these trajectories, 73,717 item clicks span 41 product categories (Appendix Figure~\ref{fig:benchmark-composition}).

For long-horizon evaluation, \textsc{UserArena} organizes 59,378 chronologically ordered sessions from 5,768 users into cross-session trajectories.
They average 10.29 sessions and 86.36 actions per user, testing whether simulators preserve preferences and behavioral tendencies across repeated visits.
Appendix Table~\ref{tab:horizon_comparison} compares their interaction horizons with related benchmarks.
For production A/B validation, \textsc{UserArena} compares the same recommendation-policy pair offline in simulation with controlled production A/B outcomes.

\subsection{Evaluation Protocol}

\label{subsec:eval}

As illustrated on the right side of \Figref{fig:eval_levels}, \textsc{UserArena} evaluates simulator fidelity at five levels of increasing scope: from whether its outputs can be parsed to whether its trajectories, personalized behavior, and session outcomes match those of real users.

We use two evaluation modes.
(1) \emph{Replay} uses the recorded state at each step of a real session to predict the logged next action.
This isolates local next-action fidelity without accumulating simulator-induced state changes.
(2) \emph{Rollout} lets the simulator select actions and determine subsequent states until the session ends.
It tests whether free-running simulated sessions reproduce real trajectories and aggregate outcomes.
We evaluate L0 under both modes, L1 under Replay, and L2--L4 under Rollout.
Appendix~\ref{app:metrics} provides the complete definitions and computation rules.

\noindent\levelbullet\textbf{L0: Environment Compliance} asks whether the environment can parse the simulator's outputs.
We report valid-action rates under Replay (RVA) and Rollout (OVA), which require each output to parse as an action executable in the current state.

\noindent\levelbullet\textbf{L1: Local Behavioral Fidelity} asks whether the simulator and real user make the same next decision from the same state.
F1 measures action-type agreement, whereas accuracy also requires the target item or comment to match when applicable.

\noindent\levelbullet\textbf{L2: Sequential Behavioral Fidelity} asks whether a free-running simulator produces realistic long-horizon dynamics.
Decision-mix Jensen--Shannon divergence (DM-JSD) compares next-decision distributions after item-page visits, capturing whether simulated and real users follow the same item-level decision patterns.
Trajectory error (TE) compares mean action count per session.

\noindent\levelbullet\textbf{L3: Personalized Population Fidelity} asks whether the simulator reproduces behavioral heterogeneity associated with users' interaction histories.
Population fidelity (PF) groups users by historical activity, purchase propensity, and category breadth, then compares simulated and real behavior within each group across session dynamics, action patterns, and product preferences.

\noindent\levelbullet\textbf{L4: Outcome Fidelity} asks whether the simulator reproduces aggregate behavioral and terminal outcomes.
Count-based click-through rate error ($\Delta$CTR) compares clicks per executed feed action, count-based add-to-cart rate error ($\Delta$ACR) compares add-to-cart actions per click, and terminal-state JSD (T-JSD) compares the distributions of final session outcomes.

\section{On-Policy Trajectory Distillation}
\label{sec:method}

\begin{figure}[t]
\centering
\includegraphics[width=\textwidth]{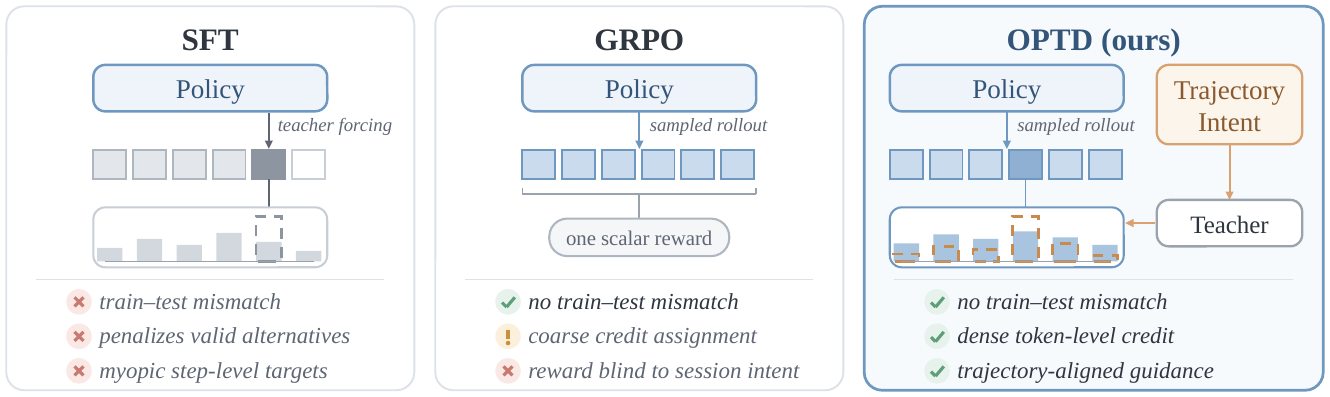}
\vspace{-10pt}
\caption{Supervision in SFT, GRPO, and \methodname. SFT learns from recorded responses, GRPO assigns a scalar reward to each generated response, and \methodname provides full-vocabulary teacher targets at every token of the student's own rollout.}
\label{fig:method}
\end{figure}

To model long-horizon shopping behavior in \textsc{UserArena}, we propose On-Policy Trajectory Distillation (\methodname).
\methodname extracts a trajectory-level intent from each complete training trajectory and provides it as privileged context to a frozen teacher.
The intent conditions dense token-level teacher supervision throughout each student response.
A per-token trust region limits how far these targets depart from predictions based only on the student's observable state.
\Figref{fig:method} contrasts the train--test mismatch and myopic targets of SFT, the coarse response-level credit assignment of GRPO, and \methodname's dense, trajectory-aligned teacher targets at every generated token.

\subsection{Task Setup}
\label{subsec:setup}

\textbf{Problem formulation.}
\textsc{UserArena} records each shopping session as a trajectory.
At every step, the user observes a page and responds to it.
Let $\tau_u$ denote the trajectory of user $u$, $x_t$ the state at step $t$, and $y^\star_t$ the user's real response at that step.
The state $x_t$ contains the current page and interaction history before the response.
The response $y^\star_t$ is a token sequence containing a short rationale followed by one structured interface action.
We denote its token at position $k$ by $y^\star_{t,k}$, its prefix before that position by $y^\star_{t,<k}$, and its token length by $|y^\star_t|$.
A user simulator is a policy $\pi_\theta(y \mid x)$ that generates response $y$ from state $x$, where $\theta$ denotes the trainable student parameters.

\textbf{On-policy sampling.} We treat $\pi_\theta$ as the student and optimize it with token-level guidance from a teacher on the student's own responses.
In contrast, standard teacher forcing on $y^\star_t$~\citep{bengio2015scheduled,ranzato2016sequence} conditions every prediction on a ground-truth prefix.
At inference, the student instead conditions on its own partial output, so errors can move it to prefixes absent from teacher-forced training~\citep{dai2021aicm,chi2022offline}.
To expose training to these self-generated prefixes, \methodname adopts on-policy distillation~\citep{agarwal2024onpolicy,zhao2026self}.
At each training step, we draw a real state $x_t$ from the empirical training distribution $\train$ and perform a one-step rollout by sampling a complete response from the current policy.
\begin{equation}
y_t = (y_{t,1}, \ldots, y_{t,|y_t|}) \sim \pi_\theta(\cdot \mid x_t),
\label{eq:onpolicy}
\end{equation}
where $y_{t,k}$ is the generated token at position $k$, $|y_t|$ is the response length, and $y_{t,<k} = (y_{t,1}, \ldots, y_{t,k-1})$ denotes its prefix before position $k$.
A teacher with privileged trajectory-level context scores each position of $y_t$, and the student learns from the resulting distributions on its own prefixes rather than only recorded prefixes.
This reduces the mismatch between teacher-forced training and autoregressive inference.

\subsection{Trajectory-Level Intent Supervision}
\label{subsec:intent}
To guide the student toward behavioral patterns that unfold across an entire session, we provide the teacher with a trajectory-level intent extracted from the complete training trajectory.

\textbf{Intent extraction.} We use a large language model as an extractor $E$ that maps the trajectory to an abstract summary of the shopper's intent,\footnote{We use Qwen3.5-397B as the extractor.}
\begin{equation}
z_u = E(\tau_u),
\label{eq:intent}
\end{equation}
where $z_u$ is a structured record of complementary typed anchors capturing trajectory-wide preferences, concerns, information needs, comparison criteria, and conditional decision tendencies.
To keep $z_u$ from serving as an answer key for individual steps, the extraction and validation rules exclude concrete actions, specific products and identifiers, exact prices, trajectory outcomes, future-page content, and the step indices supporting the summary.
The resulting intent contextualizes the current state without specifying the recorded next action.
The student never directly receives $z_u$ or observations after $x_t$; the intent influences it only through the teacher's token distributions.

\textbf{Intent-conditioned supervision.} The teacher is a frozen supervised fine-tuned checkpoint with parameters $\phi$.
Let $P_\phi$ denote its next-token distribution and let $v\in\mathcal{V}$ be a candidate token in vocabulary $\mathcal{V}$.
Given $x_t$, $z_u$, and the student prefix $y_{t,<k}$, the teacher probability of token $v$ is
\begin{equation}
q^{\mathrm{I}}_{t,k}(v) = P_\phi(v \mid x_t, z_u, y_{t,<k}),
\label{eq:qI}
\end{equation}
The student's distribution at the same position is $p^{\theta}_{t,k}(v) = \pi_\theta(v \mid x_t, y_{t,<k})$.
The intent-conditioned distribution $q^{\mathrm{I}}_{t,k}$ provides one endpoint for the regularized target constructed in \Secref{subsec:target}.
The teacher outputs are treated as constants, and no gradient reaches $\phi$.

\subsection{Trust-Region Regularization}
\label{subsec:target}
To avoid training the student to imitate predictions driven too strongly by training-only intent, we constrain how much the intent may alter the teacher's distribution at each token.
The trust region keeps intent guidance within a fixed divergence budget from a reference conditioned only on information available to the student.

\textbf{Unconditioned reference.} We construct the reference by querying the same frozen teacher on the same state and student prefix, but without $z_u$,
\begin{equation}
q^{0}_{t,k}(v) = P_\phi(v \mid x_t, y_{t,<k}).
\label{eq:q0}
\end{equation}
Because the teacher queries differ only in $z_u$, their gap isolates the effect of intent conditioning.

\textbf{Per-token trust region.} To bound that gap, we interpolate the two distributions in log-probability space, where $\lambda\in[0,1]$ is the interpolation coefficient and $v'$ ranges over the vocabulary $\mathcal{V}$:
\begin{equation}
q^{\lambda}_{t,k}(v) = \frac{\big[q^{0}_{t,k}(v)\big]^{1-\lambda}\big[q^{\mathrm{I}}_{t,k}(v)\big]^{\lambda}}{\sum_{v'\in\mathcal{V}}\big[q^{0}_{t,k}(v')\big]^{1-\lambda}\big[q^{\mathrm{I}}_{t,k}(v')\big]^{\lambda}}
= \softmax\!\Big(\log q^{0}_{t,k} + \lambda\big(\log q^{\mathrm{I}}_{t,k} - \log q^{0}_{t,k}\big)\Big),
\label{eq:geo}
\end{equation}
Setting $\lambda = 0$ recovers the unconditioned distribution and $\lambda = 1$ the fully intent-conditioned one.
Let $\lambda_{\max}\in[0,1]$ be the maximum interpolation coefficient, let $\eps\geq0$ be the divergence budget, and let $\KL(P\|Q)$ denote the Kullback--Leibler divergence from distribution $P$ to distribution $Q$.
At each position, we select the largest interpolation coefficient whose target remains within the budget:
\begin{equation}
\lambda^{\star}_{t,k} = \max\big\{\lambda \in [0, \lambda_{\max}] \;:\; \KL\big(q^{\lambda}_{t,k} \,\big\|\, q^{0}_{t,k}\big) \le \eps \big\}.
\label{eq:lambda}
\end{equation}
Along this geometric path, $\KL(q^{\lambda}_{t,k}\|q^{0}_{t,k})$ is nondecreasing in $\lambda$, making the feasible set an interval and allowing $\lambda^{\star}_{t,k}$ to be located by bisection.
The regularized target $q^{\lambda^{\star}_{t,k}}_{t,k}$ replaces $q^{\mathrm{I}}_{t,k}$ as the distribution the student matches.

The constraint adapts intent guidance at each token.
When the intent-conditioned target is feasible, the maximal permitted guidance is retained.
Otherwise, $\lambda^{\star}_{t,k}$ moves the target toward $q^{0}_{t,k}$ until it satisfies the budget.
Thus, $\eps$ bounds the token-level distributional change attributed to the intent.

\subsection{Training Objective}
\label{subsec:objective}

The \methodname objective combines a distillation term that transfers trajectory-level intent through the teacher's distributions on student responses with a grounding term that provides step-level supervision from recorded responses.
Let $\Ls_{\mathrm{dist}}$ denote the distillation loss, let $\Ls_{\mathrm{GT}}$ denote the grounding loss, let $\gamma\geq0$ be their relative weight, and let $\Ls$ denote the total loss.
The terms are computed on the same sampled states and combined as
\begin{equation}
\Ls = \Ls_{\mathrm{dist}} + \gamma\, \Ls_{\mathrm{GT}},
\label{eq:total}
\end{equation}

\textbf{Distillation term.}
Exact-match objectives credit one token sequence but no alternative phrasing or behaviorally equivalent action a real shopper might have produced~\citep{wang2025customerr1,zhang2026shopr1}.
\methodname instead matches the regularized target $q_{t,k} = q^{\lambda^{\star}_{t,k}}_{t,k}$ at every generated token using the generalized Jensen--Shannon divergence~\citep{agarwal2024onpolicy} with mixing weight $\beta\in(0,1)$ and mixture $m_{t,k} = \beta\, q_{t,k} + (1-\beta)\, p^{\theta}_{t,k}$:
\begin{equation}
\mathrm{JSD}_{\beta}\big(q_{t,k}\,\big\|\,p^{\theta}_{t,k}\big) = \beta\,\KL\big(q_{t,k}\,\big\|\,m_{t,k}\big) + (1-\beta)\,\KL\big(p^{\theta}_{t,k}\,\big\|\,m_{t,k}\big),
\label{eq:jsd}
\end{equation}
Using $\E$ for expectation over training states $x_t$ and student responses $y_t$, the distillation loss is
\begin{equation}
\Ls_{\mathrm{dist}} = \E_{x_t \sim \train,\; y_t \sim \pi_\theta(\cdot\mid x_t)}\Bigg[\frac{1}{|y_t|}\sum_{k=1}^{|y_t|} \mathrm{JSD}_{\beta}\big(q_{t,k}\,\big\|\,p^{\theta}_{t,k}\big)\Bigg].
\label{eq:ldist}
\end{equation}
The inner mean averages over generated tokens.
Because $\mathrm{JSD}_{\beta}$ is bounded, large student--target disagreement cannot produce an unbounded per-token loss.

\textbf{Grounding term.} The distillation loss conveys trajectory-level teacher guidance, whereas the recorded response $y^\star_t$ provides a step-level target at state $x_t$.
We apply a teacher-forced loss:
\begin{equation}
\Ls_{\mathrm{GT}} = \E_{(x_t,\, y^\star_t)\sim\train}\Bigg[-\frac{1}{|y^\star_t|}\sum_{k=1}^{|y^\star_t|}\log \pi_\theta\big(y^\star_{t,k} \mid x_t,\, y^\star_{t,<k}\big)\Bigg].
\label{eq:lgt}
\end{equation}
This term supervises every token of $y^\star_t$ in its original format, including the rationale, action, and terminating token.
Together, the losses combine session-wide intent with state-level behavior.

\FloatBarrier

\section{Experiments}
\label{sec:experiments}

\subsection{Experimental Setup}
\label{subsec:setup_exp}

We compare \methodname against two families of user simulators, grouped by their observation modality.
Text-based simulators observe textual page representations and include RecAgent~\citep{recagent}, Agent4Rec~\citep{agent4rec}, Shop-R1~\citep{zhang2026shopr1}, Customer-R1~\citep{wang2025customerr1}, AlignUSER~\citep{bougie2026alignuser}, and SimUSER~\citep{bougie2025simuser}.
GUI-based simulators instead observe page screenshots and include A/B Agent~\citep{zhang2026abagent} and STA~\citep{zhang2025sta}.
All methods in the main comparison use Qwen3.5-2B, enabling a controlled comparison of prompting, memory, observation modality, and post-training strategy without model-scale confounding.
Additional baseline, data-processing, implementation, and hyperparameter details are provided in Appendix~\ref{app:experimental-details}.

\subsection{Main Results}
\label{subsec:main_results}

We evaluate every simulator in two settings.
The \emph{single-session} setting evaluates each user's trajectory within one session.
The \emph{cross-session} setting evaluates a chronologically ordered trajectory spanning multiple sessions from the same user, emphasizing long-horizon behavioral prediction across visits.
We report both settings in Table~\ref{tab:main}.

\textbf{Single-session results.}
The single-session block of Table~\ref{tab:main} shows that \methodname ranks first on all eight reported L1--L4 metrics, spanning local and sequential behavioral fidelity, personalized population fidelity, and outcome fidelity.
This broad improvement aligns with its training design: trajectory-level intent guides token-level targets on the student's own rollout distribution, while the trust region and grounding term keep this guidance close to behavior supported by observable states and recorded responses.
Customer-R1's profile-conditioned interaction history is consistent with its strong local prediction, while STA's multimodal state context and reinforcement-learning-tuned checkpoint align with its strong sequential and population-level fidelity.
A/B Agent remains competitive on ACR error, but its lower Replay and Rollout valid-action rates suggest less reliable transfer to the fixed \textsc{UserArena} action protocol, limiting downstream fidelity.

\begin{table*}[t]
\caption{Single- and cross-session results on \textsc{UserArena} across L0--L4. For L1--L4, bold and underlining denote the best and second-best results, respectively, within each setting.}
\label{tab:main}
\centering
\scriptsize
\setlength{\tabcolsep}{2.8pt}
\definecolor{sessionheaderblue}{HTML}{EDF4FA}
\resizebox{\textwidth}{!}{%
\begin{tabular}{@{}clcccccccccc@{}}
\toprule
& & \multicolumn{2}{c}{\textbf{L0}} & \multicolumn{2}{c}{\textbf{L1}} & \multicolumn{2}{c}{\textbf{L2}} & \textbf{L3} & \multicolumn{3}{c}{\textbf{L4}} \\
\cmidrule(lr){3-4}\cmidrule(lr){5-6}\cmidrule(lr){7-8}\cmidrule(lr){9-9}\cmidrule(lr){10-12}
& Method & RVA $\uparrow$ & OVA $\uparrow$ & F1 $\uparrow$ & Acc. $\uparrow$ & DM-JSD $\downarrow$ & TE $\downarrow$ & PF $\uparrow$ & $\Delta$CTR $\downarrow$ & $\Delta$ACR $\downarrow$ & T-JSD $\downarrow$ \\
\midrule
\rowcolor{sessionheaderblue}
\multicolumn{12}{c}{\rule{0pt}{2.2ex}\textbf{\textit{Single-session}}} \\
\addlinespace[2pt]
\multirow{6}{*}{\rotatebox[origin=c]{90}{\textbf{Text}}}
     & RecAgent     & 0.9834 & 0.9965 & 0.1424 & 0.1507 & 0.2928 & 3.63 & 0.6941 & 0.6162 & 1.0642 & 0.1287 \\
     & Agent4Rec    & 0.9958 & 0.8192 & 0.1274 & 0.1222 & 0.2054 & 6.05 & 0.7393 & 0.4737 & 0.5086 & 0.4196 \\
     & SimUSER      & 0.9833 & 0.9651 & 0.1411 & 0.1680 & 0.1140 & 9.27 & 0.7230 & 0.2559 & 0.4897 & \underline{0.0664} \\
     & Shop-R1      & 0.9227 & 0.9551 & 0.3419 & 0.1899 & 0.1582 & 3.72 & 0.7251 & 0.0283 & 0.3623 & 0.2722 \\
     & Customer-R1 & 0.9673 & 0.9866 & \underline{0.4108} & \underline{0.3114} & 0.1954 & 2.86 & 0.7382 & \underline{0.0227} & 0.3997 & 0.2810 \\
     & AlignUSER    & 0.9902 & 0.9987 & 0.3276 & 0.2504 & 0.1973 & 6.47 & 0.6424 & 0.1049 & 0.3739 & 0.3191 \\
\midrule
\multirow{3}{*}{\rotatebox[origin=c]{90}{\textbf{GUI}}}
     & A/B Agent    & 0.4113 & 0.6302 & 0.1017 & 0.0157 & 0.6699 & 9.68 & 0.6459 & 0.0291 & 0.2039 & 0.8205 \\
     & STA          & 0.9856 & 0.9768 & 0.4030 & 0.2899 & \underline{0.0245} & \underline{1.77} & \underline{0.7958} & 0.1080 & \underline{0.1880} & 0.1519 \\
     & \methodname (\textbf{ours})  & 0.9773 & 0.9712 & \textbf{0.4178} & \textbf{0.3204} & \textbf{0.0156} & \textbf{1.72} & \textbf{0.8692} & \textbf{0.0026} & \textbf{0.0430} & \textbf{0.0387} \\
\midrule
\rowcolor{sessionheaderblue}
\multicolumn{12}{c}{\rule{0pt}{2.2ex}\textbf{\textit{Cross-session}}} \\
\addlinespace[2pt]
\multirow{6}{*}{\rotatebox[origin=c]{90}{\textbf{Text}}}
     & RecAgent     & 0.9930 & 0.9961 & 0.1032 & 0.1747 & 0.2157 & 7.66 & 0.6621 & 0.5521 & 0.7428 & 0.1268 \\
     & Agent4Rec    & 0.9943 & 0.8430 & 0.1294 & 0.1455 & 0.1541 & 1.57 & 0.7657 & 0.4098 & 1.1625 & 0.4424 \\
     & SimUSER      & 0.9860 & 0.9628 & 0.0946 & 0.1629 & 0.0387 & 3.51 & 0.7492 & 0.1872 & 0.3502 & 0.0275 \\
     & Shop-R1      & 0.9476 & 0.9920 & \underline{0.3533} & 0.1975 & 0.1310 & 1.35 & 0.8739 & 0.0905 & 0.3514 & 0.0941 \\
     & Customer-R1 & 0.9531 & 0.9871 & 0.3415 & 0.1902 & 0.0574 & 1.91 & 0.8628 & 0.0407 & 0.1515 & 0.0244 \\
     & AlignUSER    & 0.9545 & 0.9864 & 0.3172 & 0.1933 & \underline{0.0370} & 2.11 & 0.8732 & 0.0485 & 0.1461 & \underline{0.0099} \\
\midrule
\multirow{3}{*}{\rotatebox[origin=c]{90}{\textbf{GUI}}}
     & A/B Agent    & 0.5182 & 0.5777 & 0.1094 & 0.0265 & 0.5709 & 4.17 & 0.6917 & 0.0615 & \textbf{0.0096} & 0.7980 \\
     & STA          & 0.9563 & 0.9830 & \textbf{0.3556} & \underline{0.2919} & 0.0465 & \underline{0.81} & \underline{0.8911} & \textbf{0.0099} & 0.1577 & 0.0173 \\
     & \methodname (\textbf{ours})  & 0.9578 & 0.9792 & 0.3456 & \textbf{0.2969} & \textbf{0.0284} & \textbf{0.72} & \textbf{0.8929} & \underline{0.0357} & \underline{0.1399} & \textbf{0.0077} \\
\bottomrule
\end{tabular}
}
\end{table*}

\textbf{Cross-session results.}
The cross-session block of Table~\ref{tab:main} uses the same summary metrics.
Among all methods, \methodname ranks first on five of the eight L1--L4 metrics and second on two more.
It achieves the strongest L2 sequential behavioral fidelity, with the lowest decision-mix divergence and trajectory-length error, together with the highest L3 personalized population fidelity.
These results demonstrate \methodname's gains in preserving long-horizon trajectory structure and personalized behavioral patterns across sessions.
Its leading L1 accuracy separately indicates the strongest step-wise action prediction under Replay.
STA leads macro-F1 and CTR error, suggesting that its reinforcement-learning post-training benefits selected local and outcome-level metrics, although these gains are not consistent across fidelity levels.

\newpage
\subsection{Production A/B Validation}
\label{subsec:ab-validation}

\begin{wraptable}{r}{0.48\textwidth}
\vspace{-\intextsep}
\vspace{-\baselineskip}
\vspace{2pt}
\setlength{\belowcaptionskip}{4pt}
\caption{Production A/B validation of the same ranking-policy pair.}
\label{tab:abtest-ctr}
\centering
\small
\setlength{\tabcolsep}{9pt}
\begin{tabular}{@{}lrr@{}}
\toprule
Simulator & $\Delta$PV-CTR ratio & Effect error $\downarrow$ \\
\midrule
Shop-R1 & 0.162$\times$ & 0.838 \\
Customer-R1 & 0.191$\times$ & 0.809 \\
AlignUSER & 0.020$\times$ & 0.980 \\
STA & 0.297$\times$ & 0.703 \\
\methodname & \textbf{0.713$\times$} & \textbf{0.287} \\
Qwen3.5-2B & 0.004$\times$ & 0.996 \\
\midrule
GPT-6-Sol & 1.576$\times$ & 0.576 \\
GPT-6-Luna & 2.154$\times$ & 1.154 \\
\bottomrule
\end{tabular}
\vspace{-4pt}
\end{wraptable}

To assess whether simulator-based policy comparisons reflect real-user responses to recommendation changes, we conduct an offline validation against outcomes recorded for the control and treatment arms of a production A/B test.
We evaluate the same pair of ranking policies in each simulator and compare the simulated policy effect with the recorded production effect.
In simulation, we assign a frozen pool of 10,000 users evenly to the two policy arms and hold the users and product catalog fixed across all simulators.
Because the ranking policies alter item exposure, we measure their effect using the between-arm difference in page-view click-through rate ($\Delta$PV-CTR).
The effect ratio divides the simulated $\Delta$PV-CTR by its online value, and the effect error is the absolute deviation of this ratio from 1.
Among the evaluated user simulators, \methodname most closely reproduces the online policy effect, reaching an effect ratio of 0.713$\times$.
GPT-6-Sol and GPT-6-Luna instead overestimate the online response, indicating that general-purpose model capacity does not directly provide behavioral calibration in this setting.
Full experimental details appear in Appendix~\ref{app:ab-validation}.
\WFclear

\subsection{In-Depth Analyses}
\label{subsec:in-depth}

We further examine \methodname through component ablations and a Turing-style human evaluation.
Appendix~\ref{app:additional-analyses} provides an efficiency analysis and qualitative case studies.

\begin{figure*}[t]
    \centering
    \begin{minipage}[b]{0.64\textwidth}
        \centering
        \includegraphics[width=\linewidth]{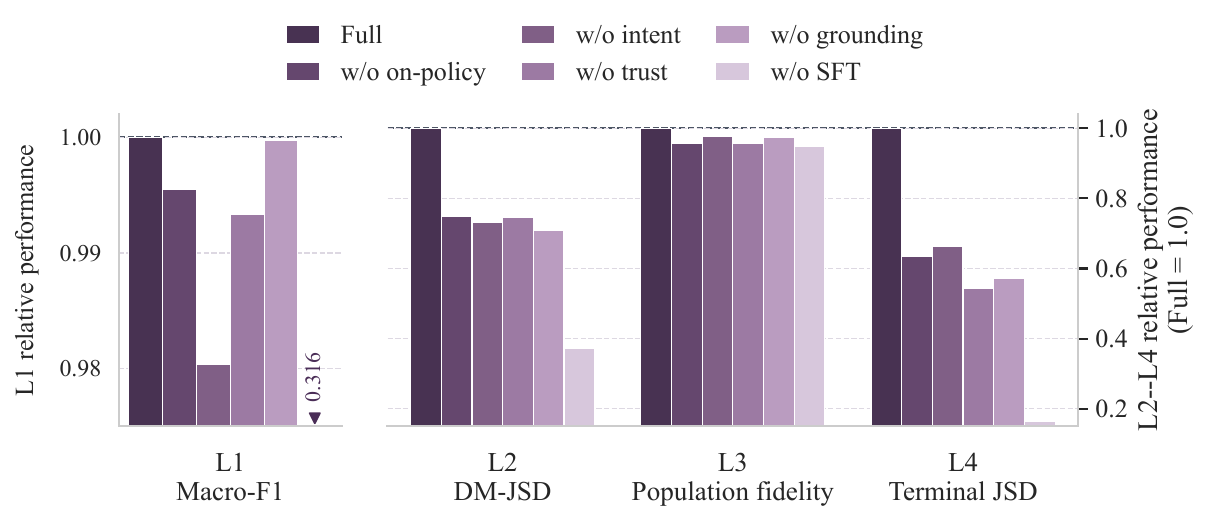}
        \small\textbf{(a)} Relative performance by fidelity level.
    \end{minipage}
    \hfill
    \begin{minipage}[b]{0.32\textwidth}
        \centering
        \includegraphics[width=\linewidth]{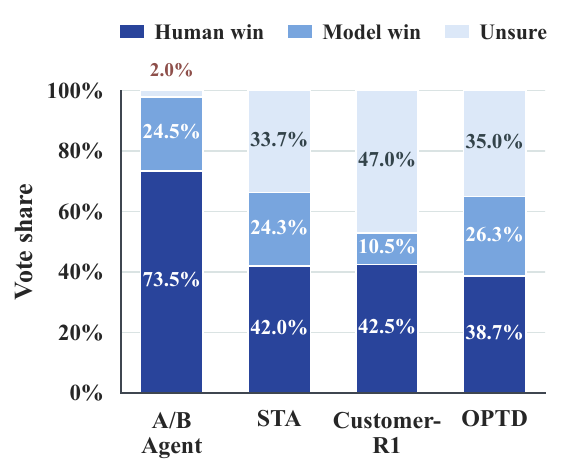}
        \small\textbf{(b)} Turing-style human evaluation.
    \end{minipage}
    \caption{Ablation and human-evaluation results. (a) Component performance relative to the full model. (b) Human win denotes correct identification of the simulated trajectory, Model win denotes misidentification of the real trajectory as simulated, and Unsure denotes abstention.}
    \label{fig:ablation-turing}
\end{figure*}

\textbf{Ablation study.}
To understand the contribution of each component within \methodname, we construct five ablations with all other settings fixed.
(1) \emph{w/o on-policy} replaces student-generated responses with logged responses, so distillation operates on recorded prefixes rather than the student's own prefixes.
(2) \emph{w/o intent} removes trajectory-level intent from the teacher context.
(3) \emph{w/o trust region} uses the fully intent-conditioned teacher distribution without trust-region interpolation.
(4) \emph{w/o grounding} sets the grounding weight to zero.
(5) \emph{w/o SFT} initializes the teacher and student from the base model.
Figure~\ref{fig:ablation-turing}(a) reports one representative metric from each behavioral fidelity level L1--L4.
First, removing any component reduces performance on at least one selected metric, showing that each contributes to behavioral fidelity.
Second, the \emph{w/o SFT} variant degrades most markedly across the selected metrics, indicating that the warm-up establishes the instruction-following and action-generation abilities required for subsequent trajectory distillation.

\textbf{Turing-style human evaluation.}
Automatic metrics quantify agreement with recorded behavior, but they do not directly test whether a complete simulated trajectory appears human to an observer.
We therefore define a source-blinded Turing-style evaluation~\citep{turing1950computing} that compares \methodname with the text-based Customer-R1~\citep{wang2025customerr1} and two GUI-based simulators, STA~\citep{zhang2025sta} and A/B Agent~\citep{zhang2026abagent}.
For each user context, three annotators independently compare an anonymized, order-randomized real--simulated pair and identify the machine-generated trajectory or choose \emph{Unsure}.
Each simulator is evaluated on 200 trajectory pairs sampled from the held-out test set.
Figure~\ref{fig:ablation-turing}(b) shows that \methodname has the lowest Human win rate and the highest Model win rate among the four simulators.
Appendix~\ref{app:turing-interface} provides the complete protocol and experiment analysis.

\section{Related Work}
\label{sec:related_work}

\textbf{Interactive recommendation benchmarks.} Recommendation benchmarks have evolved from static preference records toward richer behavioral data and interactive evaluation. MovieLens~\citep{Harper2015MovieLens} records movie ratings, Amazon Reviews~\citep{Hou2026Bridging} pairs product ratings with review text, Qilin~\citep{Chen2025Qilin} adds multimodal app-level sessions, and RL4RS~\citep{rl4rs} provides real-world sequential interactions. Agent-oriented benchmarks further broaden the task interface: RecBench+~\citep{Huang2026RecBenchPlus} evaluates personalized recommendation through natural-language requests, AgentRecBench~\citep{agentrecbench} introduces interactive recommendation scenarios, and OPeRA~\citep{opera} and OmniBehavior~\citep{omnibehavior} provide richer evidence for human behavior simulation. Simulation platforms such as RecSim~\citep{recsim}, Virtual-Taobao~\citep{virtualtaobao}, and KuaiSim~\citep{kuaisim} further support interactive evaluation across temporal scales. Yet existing resources rarely record, throughout long-horizon sessions, what real users see and how they act, or test whether simulator-based comparisons reproduce policy effects observed in production A/B tests. UserArena provides both forms of evidence through step-aligned real-world shopping trajectories and offline validation against outcomes from a controlled production A/B test.

\textbf{User simulator development.} Early systems represented users through configurable or parameterized dynamics, including RecSim~\citep{recsim}, RecoGym~\citep{recogym}, and the probabilistic RecSim NG framework~\citep{recsimng}. Log-driven approaches then learned interaction dynamics from observed behavior: Virtual-Taobao~\citep{virtualtaobao} combines generative customer modeling with adversarial imitation, while the Generative Adversarial User Model~\citep{ganusermodel} learns user dynamics and rewards for model-based policy training. IRecGAN~\citep{irecgan} incorporates generated trajectories into adversarial user-model learning. More recent LLM-based simulators, including RecAgent~\citep{recagent}, Agent4Rec~\citep{agent4rec}, and SUBER~\citep{suber}, increase the expressiveness of profile-conditioned behavior, while Customer-R1~\citep{wang2025customerr1}, Shop-R1~\citep{zhang2026shopr1}, and RecVerse~\citep{tang2026towards} apply reinforcement learning to shopping simulation. Despite this progress, two optimization challenges remain. \textbf{(1) Sparse rewards:} most methods use exact-match rewards, treating plausible alternatives and invalid behaviors alike. \textbf{(2) Exposure bias:} teacher forcing trains only on ground-truth prefixes, leaving models unprepared for self-generated prefixes and vulnerable to compounding errors at inference. \methodname addresses both through dense distributional supervision on student-sampled responses and trajectory-level intent for long-horizon behavioral guidance.

\section{Conclusion}
\label{sec:conclusion}

We introduced \textsc{UserArena}, an interactive benchmark combining a production-grounded shopping environment with fine-grained, long-horizon real-user trajectories.
We also developed a five-level protocol that evaluates simulator quality from environment compliance and local decisions to sequential, personalized population, and outcome fidelity.
For this long-horizon setting, we proposed On-Policy Trajectory Distillation (\methodname), which uses trajectory-level intent to guide dense teacher supervision on the student's own rollouts.
\methodname achieves the strongest overall fidelity under this protocol and most closely reproduces the production policy effect in the controlled comparison.

As future work we will develop a leaderboard platform around \textsc{UserArena} to support standardized and continuously updated evaluation of user simulators. 
We will investigate stronger long-horizon simulators that preserve evolving preferences and behavioral consistency across sessions.

\section*{AI Use Statement}

We used generative AI tools to edit the manuscript for clarity and readability.
We also used them to assist with translating Chinese interface text and with translating and shortening Chinese item titles for presentation in figures and qualitative examples.
These presentation edits did not modify the underlying trajectories, annotations, evaluation labels, or experimental results.
The authors manually reviewed all AI-assisted text and translations against the original Chinese materials and take responsibility for the final content of the paper.

\section*{Reproducibility Statement}

We support reproducibility by describing the benchmark construction and evaluation protocol in Section~\ref{sec:benchmark}, the complete \methodname formulation in Section~\ref{sec:method}, and the experimental setup in Section~\ref{sec:experiments}.
Appendix~\ref{app:benchmark-details} provides dataset statistics and metric definitions, while Appendix~\ref{app:experimental-details} documents data processing, baseline configurations, implementation details, and prompt structure.
Additional analyses are reported in Appendix~\ref{app:additional-analyses}.

\section*{Ethics and Data Statement}

The dataset was derived from de-identified interaction data collected during normal platform operation. 
Its use for research was authorized in accordance with the platform's privacy policy and internal data-governance procedures. 
We selected sessions and their associated item pools from user behavior logs and applied data cleaning and de-identification before release. 
Product metadata contain only product titles and English top-level category labels. We removed information that could directly identify users and replaced production identifiers with non-linkable sequential identifiers.

The released interface images contain no GUI screenshots captured from real users. 
We reconstruct the two-column feed and item-detail pages in an offline React environment using the de-identified sessions and item pools, and then batch-render the resulting interfaces. 
The dataset is released for non-commercial research under a data use agreement.
The agreement prohibits attempts to re-identify users, linkage with external personal data, and unauthorized redistribution.

\section*{Acknowledgments}

This work was supported by Alibaba Group through the Alibaba Innovative Research Program.

\bibliography{iclr2027_conference}
\bibliographystyle{iclr2027_conference}

\appendix
\clearpage
\section{Additional Benchmark Details}
\label{app:benchmark-details}

\subsection{Comparison of Interaction Horizons}
\label{app:horizon}

Table~\ref{tab:benchmark_comparison} provides a qualitative comparison of benchmark capabilities. 
To further substantiate its long-horizon dimension, Table~\ref{tab:horizon_comparison} quantifies the interaction horizons of real-user shopping benchmarks with a full action space. 
OPeRA~\citep{opera} and USB~\citep{tang2026towards} evaluate behavior within individual shopping sessions, whereas \textsc{UserArena} captures longitudinal user trajectories spanning multiple natural and consecutive shopping sessions. 
A/B Agent is excluded because its interaction trajectories are synthetically generated rather than collected from real users.

\begin{table}[!ht]
\caption{Interaction-horizon statistics for real-user shopping benchmarks with a full action space. For UserArena, statistics are reported for the cross-session subset.}
\label{tab:horizon_comparison}
\centering
\small
\setlength{\tabcolsep}{5pt}
\begin{tabular}{@{}lrrlrr@{}}
\toprule
Benchmark & \# Users & \# Sessions & Trajectory Scope & Avg. Sessions & Avg. Actions \\
\midrule
OPeRA & 51 & 527 & Single-session & 1.00 & 11.11 \\
USB & 5,222 & 5,274 & Single-session & 1.00 & 13.24 \\
\textbf{\textsc{UserArena}} & \textbf{5,768} & \textbf{59,378} & \textbf{Cross-session} & \textbf{10.29} & \textbf{86.36} \\
\bottomrule
\end{tabular}
\end{table}

\subsection{Dataset and Behavior Statistics}
\label{app:benchmark-statistics}

We provide additional analyses of \textsc{UserArena}'s behavioral composition, interaction horizons, and coverage.

\begin{figure*}[t]
  \centering
  \begin{minipage}[b]{0.64\textwidth}
    \centering
    \includegraphics[width=\linewidth]{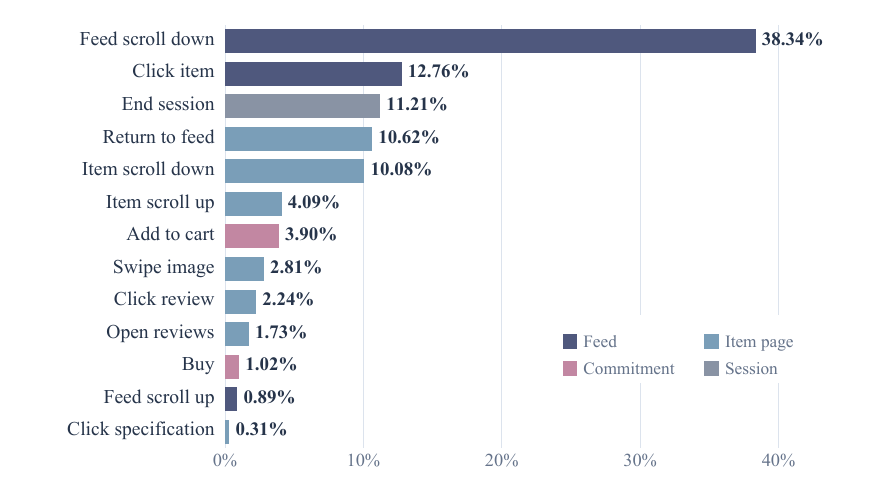}
    \small\textbf{(a) Decision-action distribution.}
  \end{minipage}
  \hfill
  \begin{minipage}[b]{0.32\textwidth}
    \centering
    \includegraphics[width=\linewidth]{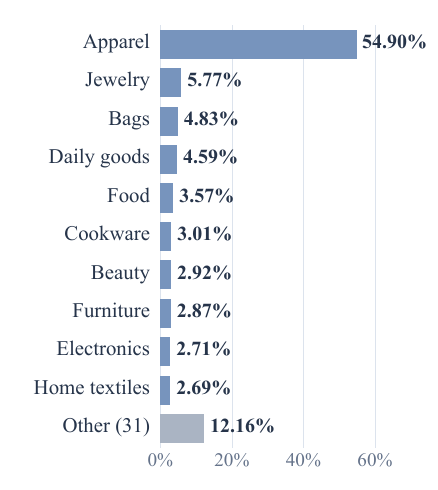}
    \small\textbf{(b) Clicked item categories.}
  \end{minipage}
  \caption{Behavioral composition of \textsc{UserArena}. (a) Distribution of decision actions, grouped into feed, item-page, commitment, and session actions. (b) Distribution of clicked items across product categories, with Other aggregating the remaining 31 categories.}
  \label{fig:benchmark-composition}
\end{figure*}

\textbf{Action and item composition.}
Figure~\ref{fig:benchmark-composition}(a) reports the proportion of every action type in \textsc{UserArena}.
Feed-page downward scrolling is the most frequent action at 38.34\%, followed by item clicks at 12.76\%, session termination at 11.21\%, returns to the feed at 10.62\%, and item-page downward scrolling at 10.08\%.
Add-to-cart and purchase actions account for 3.90\% and 1.02\% of decisions, respectively.
This imbalance supports using macro-averaged local metrics and distributional trajectory metrics rather than allowing frequent navigation actions to dominate evaluation.
Figure~\ref{fig:benchmark-composition}(b) shows that the 73,717 item clicks span 41 product categories.
Apparel accounts for 54.90\% of clicks, while the remaining 45.10\% covers 40 categories, including jewelry, bags, daily goods, food, cookware, beauty, furniture, electronics, and home textiles.

\begin{figure*}[t]
  \centering
\includegraphics[
  width=\textwidth,
  trim=0 36bp 0 0,
  clip
]{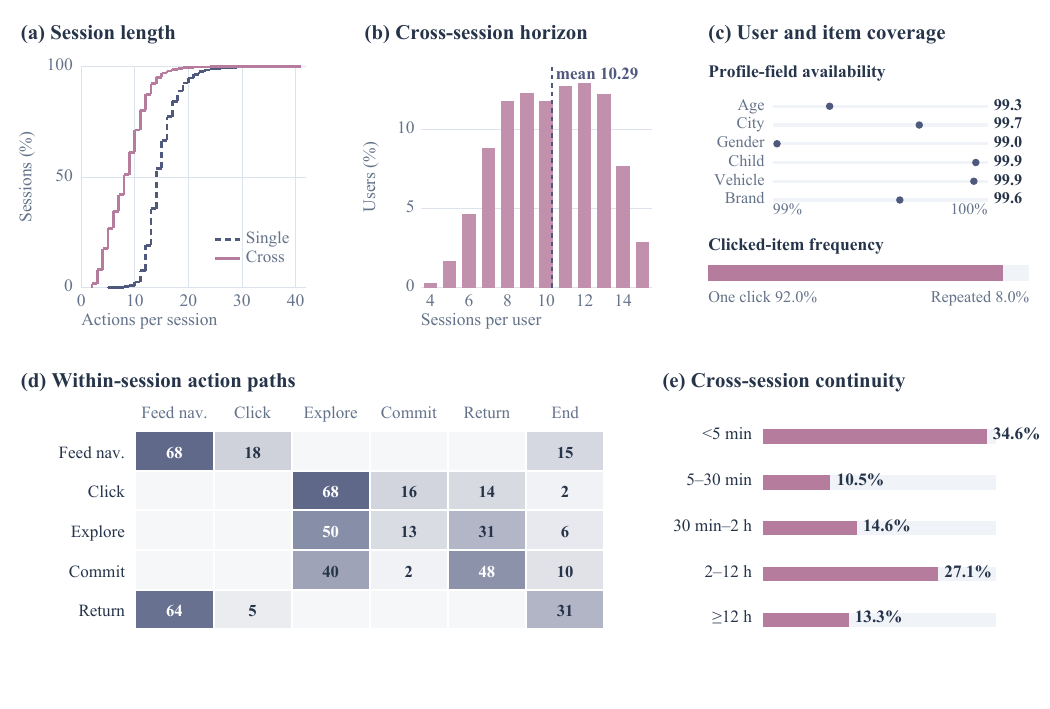}
  \caption{Additional statistics for \textsc{UserArena}. (a) Empirical cumulative distributions of session length in the single- and cross-session settings. (b) Distribution of sessions per cross-session user. (c) User-profile coverage and click frequency across distinct items. (d) Row-normalized within-session transitions between semantic action groups. (e) Time gaps between adjacent cross-session pairs, excluding overlapping sessions.}
  \label{fig:benchmark-statistics}
\end{figure*}

\textbf{Session and user horizons.}
Figure~\ref{fig:benchmark-statistics}(a) shows that individual sessions contain a median of 14 actions in the single-session setting and 8 actions in the cross-session setting.
Cross-session user trajectories become substantially longer by linking repeated visits: each user contributes 4--15 chronologically ordered sessions, with 10.29 sessions and 86.36 actions on average (Figure~\ref{fig:benchmark-statistics}(b)).

\textbf{User and item coverage.}
Figure~\ref{fig:benchmark-statistics}(c) shows that availability across the six user-profile fields ranges from 99.0\% to 99.9\%, while 92.0\% of the 67,267 clicked items appear only once.
The trajectories cover 566,102 products overall, and the most-clicked 1\% of items account for only 3.23\% of clicks, indicating that interactions extend well beyond a small set of repeatedly observed products.

\textbf{Behavioral paths.}
Figure~\ref{fig:benchmark-statistics}(d) aggregates adjacent actions into six semantic groups and reports row-normalized transition probabilities.
The resulting paths expose repeated feed navigation, item exploration after item clicks, and returns to the feed after item-page decisions.
These transitions capture multi-step shopping processes that cannot be represented by isolated action frequencies.

\textbf{Cross-session continuity.}
The cross-session subset yields 53,610 adjacent session pairs from the same users.
As shown in Figure~\ref{fig:benchmark-statistics}(e), successive sessions span intervals from minutes to more than 12 hours, and 57.8\% revisit a product category present in the preceding session.
The benchmark therefore captures both rapid returns and longer-term revisits while preserving each user's chronological session order.

\subsection{Evaluation Metric Details}
\label{app:metrics}
Section~\ref{sec:benchmark} introduces the five-level evaluation protocol and its metrics.
Here we specify how each metric reported in Table~\ref{tab:main} is computed.
We first establish notation shared across the definitions below.
The symbol $\mathcal{R}$ denotes logged human sessions and $\mathcal{S}$ denotes simulator-generated sessions.
Replay predicts one next action at each logged human state, while rollout generates complete sessions.
For a finite set $Z$, $|Z|$ is its number of elements.

\subsubsection{L0: Environment Compliance (Replay and Rollout)}
\textbf{Replay and Rollout valid-action rates (RVA and OVA).}
Let mode $m\in\{\mathrm{R},\mathrm{O}\}$ denote Replay or Rollout, and let $N_m$ denote its number of model outputs.
For output $j$, $V_{m,j}$ denotes 1 when the output parses into an action that can execute in the observed state, including successful target resolution, and 0 otherwise.
Then
\begin{equation}
\operatorname{VA}_m=\frac{1}{N_m}\sum_{j=1}^{N_m}V_{m,j}.
\end{equation}
RVA is $\operatorname{VA}_{\mathrm R}$ and OVA is $\operatorname{VA}_{\mathrm O}$.
Every output remains in the denominator.
An unparseable output or fallback substitution has $V_{m,j}=0$.

\subsubsection{L1: Local Behavioral Fidelity (Replay)}

\textbf{Macro-F1 (F1).}
At replay step $i$, $y_i$ is the logged action type and $\hat y_i$ is the predicted type.
An unparseable or invalid prediction is an incorrect empty prediction.
Let $\mathcal{C}$ be the union of logged and valid predicted types.
For type $c\in\mathcal{C}$, $\mathrm{TP}_c$ counts $y_i=c,\hat y_i=c$, $\mathrm{FP}_c$ counts $y_i\neq c,\hat y_i=c$, and $\mathrm{FN}_c$ counts $y_i=c,\hat y_i\neq c$.
We compute
\begin{align}
\mathrm{Precision}_c &= \frac{\mathrm{TP}_c}{\mathrm{TP}_c+\mathrm{FP}_c},\\
\mathrm{Recall}_c &= \frac{\mathrm{TP}_c}{\mathrm{TP}_c+\mathrm{FN}_c},\\
F_{1,c} &= \frac{2\,\mathrm{Precision}_c\,\mathrm{Recall}_c}
{\mathrm{Precision}_c+\mathrm{Recall}_c},\\
\operatorname{F1} &= \frac{1}{|\mathcal{C}|}\sum_{c\in\mathcal{C}}F_{1,c}.
\end{align}
A zero denominator gives zero, and the final mean weights action types equally.

\textbf{Accuracy (Acc.).}
An exact match requires the action type to match.
Item and comment clicks must also match their targets, using an exact item-title match only when comparable item identifiers are unavailable.
Missing targets and invalid predictions are misses.
Let $\mathcal{C}_{\mathrm{NS}}$ denote the logged action types after excluding feed-page and item-page scroll-up and scroll-down actions.
For type $c\in\mathcal{C}_{\mathrm{NS}}$, $n_c$ denotes the number of logged steps and $h_c$ denotes the exact matches.
Then
\begin{equation}
\operatorname{Acc.}=\frac{1}{|\mathcal{C}_{\mathrm{NS}}|}
\sum_{c\in\mathcal{C}_{\mathrm{NS}}}\frac{h_c}{n_c}.
\end{equation}
This weights every retained action type equally.
If $\mathcal{C}_{\mathrm{NS}}$ is empty, Acc. is not measured.
We exclude the four scroll actions because they are target-free navigation primitives whose exact-match criterion reduces to action-type agreement.
Treating them as four separate exact-match classes would overrepresent low-level navigation and duplicate the action-type signal already measured by F1, rather than testing decisions with behavioral or target-specific consequences.

\subsubsection{L2: Sequential Behavioral Fidelity (Rollout)}

\textbf{Decision-mix JSD (DM-JSD).}
For every item-page visit, we record the first decision in $\mathcal{D}=\{\text{return},\text{cart},\text{buy},\text{end},\text{open}\}$.
Let $q_X(d)$ denote the fraction of visits in session set $X\in\{\mathcal{R},\mathcal{S}\}$ whose decision is $d\in\mathcal{D}$.
For distributions $P$ and $Q$ over $\mathcal{D}$, let $M(d)=(P(d)+Q(d))/2$.
We use the base-2 Jensen--Shannon divergence
\begin{equation}
\operatorname{JSD}(P,Q)=\frac{1}{2}\sum_{d\in\mathcal{D}}P(d)\log_2\frac{P(d)}{M(d)}
+\frac{1}{2}\sum_{d\in\mathcal{D}}Q(d)\log_2\frac{Q(d)}{M(d)},
\end{equation}
where a zero-probability term contributes zero.
JSD ranges from 0 for identical distributions to 1 for disjoint distributions.
If both session sets contain no item-page visits, DM-JSD is 0. If exactly one set contains no visits, DM-JSD is 1.
We compute
\begin{equation}
\operatorname{DM\text{-}JSD}=\operatorname{JSD}(q_{\mathcal R},q_{\mathcal S}).
\end{equation}

\textbf{Trajectory error (TE).}
For session $s$, $\ell_s$ denotes the number of executed policy actions, including the terminal action when present.
For session set $X\in\{\mathcal{R},\mathcal{S}\}$, $\bar\ell_X=|X|^{-1}\sum_{s\in X}\ell_s$ denotes mean trajectory length.
Truncated sessions contribute their observed executed length.
We compute
\begin{equation}
\operatorname{TE}=|\bar\ell_{\mathcal S}-\bar\ell_{\mathcal R}|.
\end{equation}

\subsubsection{L3: Personalized Population Fidelity (Rollout)}

\textbf{Population fidelity (PF).}
PF tests whether a simulator preserves behavioral differences among shopper groups rather than reproducing only population-average behavior.
We compute it in four steps.

\textit{Step 1: Construct user cohorts.}
Using only history recorded before the evaluated session, we group users by activity, purchase propensity, and category breadth.
We fit cohort boundaries on $\mathcal{R}$ and apply them unchanged to $\mathcal{S}$.

\textit{Step 2: Summarize cohort behavior.}
For every cohort in each session set, we construct a behavioral signature with eight blocks: session length, action distribution, information-depth distribution, item-page entry, item-page return, conversion, item category, and item price.
Information depth counts information-gathering actions before the first return, cart, purchase, or session end within an item-page visit.
We bin depth into four real-data quantiles and price into five, and represent categories using the ten most frequent real categories plus an other category.
Item-page entry is the fraction of sessions that visit an item page, while item-page return is the mean fraction of item-page visits that return to the feed among sessions containing such a visit.

\textit{Step 3: Compare real and simulated cohorts.}
For cohort $g$, let $d_{g,b}\in[0,1]$ denote the real--simulated discrepancy for block $b$.
We use the base-2 JSD defined above for distribution blocks and absolute differences for rate blocks.
The conversion discrepancy averages the cart-rate and purchase-rate gaps, while the session-length gap is scaled by the population standard deviation in $\mathcal{R}$ and capped at 1.
If that standard deviation is zero, we scale by the real mean session length, or by 1 when both quantities are zero.
The fidelity of cohort $g$ is
\begin{equation}
F_g=1-\frac{1}{8}\sum_{b=1}^{8}d_{g,b}.
\end{equation}

\textit{Step 4: Aggregate across cohorts and schemes.}
For scheme $k$, $\mathcal{G}_k$ contains cohorts with at least 30 real and 30 simulated sessions, and $F_k=|\mathcal{G}_k|^{-1}\sum_{g\in\mathcal{G}_k}F_g$.
Let $\mathcal{K}$ contain schemes with at least two such cohorts.
Then
\begin{equation}
\operatorname{PF}=\frac{1}{|\mathcal{K}|}\sum_{k\in\mathcal{K}}F_k.
\end{equation}
Both averaging stages use equal weights.
If $\mathcal{K}$ is empty, PF is not measured.

\subsubsection{L4: Outcome Fidelity (Rollout)}

\textbf{Count-based CTR error ($\Delta$CTR).}
For session set $X\in\{\mathcal{R},\mathcal{S}\}$, $c_X$ denotes the total number of click actions and $f_X$ denotes the total number of executed feed-page actions.
The count CTR is $v_{\mathrm{click}}^X=c_X/f_X$.
We compute
\begin{equation}
\Delta\operatorname{CTR}=|v_{\mathrm{click}}^{\mathcal S}-v_{\mathrm{click}}^{\mathcal R}|.
\end{equation}
If $f_X=0$, we set $v_{\mathrm{click}}^X=0$.

\textbf{Count-based ACR error ($\Delta$ACR).}
For session set $X\in\{\mathcal{R},\mathcal{S}\}$, $a_X$ denotes the total number of add-to-cart actions and $c_X$ denotes the total number of click actions.
The count add-to-cart rate is $v_{\mathrm{cart}}^X=a_X/c_X$.
We compute
\begin{equation}
\Delta\operatorname{ACR}=|v_{\mathrm{cart}}^{\mathcal S}-v_{\mathrm{cart}}^{\mathcal R}|.
\end{equation}
If $c_X=0$, we set $v_{\mathrm{cart}}^X=0$.

\textbf{Terminal-state JSD (T-JSD).}
Each session receives one outcome: purchase, cart only, browse only, or truncated.
Purchase has first priority, followed by a truncation without normal termination, followed by cart, and then browse only.
For session set $X\in\{\mathcal{R},\mathcal{S}\}$ and outcome $o$, let $n_X(o)$ be the number of sessions assigned to $o$.
We define the empirical terminal-outcome distribution as $T_X(o)=n_X(o)/|X|$ over the four outcomes.
Then
\begin{equation}
\operatorname{T-JSD}=\operatorname{JSD}(T_{\mathcal R},T_{\mathcal S}).
\end{equation}

\section{Experiment Setup Details}
\label{app:experimental-details}

\subsection{Baseline Details}
\label{app:baseline-details}

We compare against eight representative user simulators while preserving each method's observation modality, context construction, memory mechanism, and training objective.
All methods use the common \textsc{UserArena} environment, action space, source test split, and output schema.

\begin{enumerate}[leftmargin=*,itemsep=2pt,topsep=3pt]
    \item \textbf{RecAgent}~\citep{recagent} conditions an LLM agent on a user profile and hierarchical sensory, short-term, and long-term memories. The memories are updated only with interactions that have already occurred.
    \item \textbf{Agent4Rec}~\citep{agent4rec} represents a user through profile, memory, and reflection modules. Retrieved memories and a compact reflection on prior interactions inform each action.
    \item \textbf{SimUSER}~\citep{bougie2025simuser} constructs a profile from pre-session history and updates category preferences from executed interactions. Our adapted version retains one profile update per session and one action prediction per decision.
    \item \textbf{Shop-R1}~\citep{zhang2026shopr1} is a text-based simulator trained with supervised fine-tuning followed by reinforcement learning. It receives the textual page state and interaction history but not the user profile.
    \item \textbf{Customer-R1}~\citep{wang2025customerr1} extends the Shop-R1 training recipe with the shopper profile, recent clicks and purchases, and the executed interaction history.
    \item \textbf{AlignUSER}~\citep{bougie2026alignuser} augments supervised training with counterfactual actions and teacher-generated comparison rationales. Its evaluation prompt contains only the profile, past interactions, and current state.
    \item \textbf{A/B Agent}~\citep{zhang2026abagent} is a GUI-based simulator that combines the current screenshot with a user profile, retrieved long-term memory, working memory, and a fatigue mechanism.
    \item \textbf{STA}~\citep{zhang2025sta} is a GUI-based simulator that predicts a rationale and structured action from the current screenshot, textual page state, and interaction history. We use its reinforcement-learning-tuned checkpoint.
\end{enumerate}

\subsection{Data Processing and Implementation}
\label{app:implementation-details}

\textbf{Data processing.}
We normalize the raw interaction logs into the 13-action \textsc{UserArena} space and reconstruct the state immediately preceding each recorded decision.
The visual observation and textual page state are drawn from the same pre-action interface state.
We stratify users into training, validation, and test sets at an 8:1:1 ratio using seed 42, assigning all sessions from each user to the same split for both single- and cross-session experiments.
Training and validation examples may be resampled to balance action coverage, whereas every test trajectory is retained once and is never resampled.
Replay reconstructs each recorded state independently, while Rollout starts from the same initial context and thereafter advances only with actions accepted by the environment.

\textbf{Implementation details.}
To control for backbone capacity, all methods in the main comparison, including the prompting-only baselines, use Qwen3.5-2B~\citep{qwen2026qwen35} with a maximum context length of 32k tokens.
This shared-backbone setting isolates differences in prompting, memory, observation modality, and post-training strategy from differences in model scale.
We standardize the interaction interface of every trainable baseline and \methodname.
Their base prompts use the same task instruction, action definitions, and response schema, while retaining any method-specific context described in Appendix~\ref{app:baseline-details}.
We use Qwen3.5-397B~\citep{qwen2026qwen35} to generate a short first-person rationale for each recorded action from the corresponding user context, interaction history, and current state.
Each SFT target pairs this rationale with the recorded structured action.
The rationale encourages the model to summarize relevant evidence from the user context, interaction history, and current state before selecting an action, and using this format for every trainable method keeps the supervision interface consistent.
We cap each GUI image at 200,704 input pixels.
For \methodname, we perform an SFT warm-up on the same interaction format so that the resulting checkpoint can follow task instructions and generate a short rationale followed by a valid action from the current state.
We then freeze this checkpoint as the teacher and initialize the student from it.
We use ROLL~\citep{wang2025roll} for distributed training, vLLM~\citep{kwon2023vllm} for on-policy generation, and Megatron-Core~\citep{shoeybi2019megatron} for student optimization and teacher inference.
\methodname training uses BF16 precision, tensor and sequence parallelism, distributed optimization, and activation recomputation to reduce GPU memory use.

Replay uses greedy decoding and Rollout samples with temperature 1.0, top-$p$ 0.95.
Generation is capped at 512 tokens per decision.
We cap Rollout at 31 executed actions for single-session evaluation and 42 for cross-session evaluation, one more than the maximum recorded action count in each setting.
A trajectory is marked as truncated when it reaches the corresponding limit without terminating.

\subsection{Training Hyperparameters}
\label{app:training-hyperparameters}

All trainable baselines and \methodname use a peak learning rate of $2 \times 10^{-8}$.
For GRPO-trained baselines, each rollout batch contains 64 samples organized into groups of eight responses per prompt.
We use an initial KL-controller coefficient of 0.001 and a KL-loss coefficient of 0.01, with temperature 0.99 and top-$p$ 0.95 for rollout generation.
For \methodname, we optimize the student with Adam and zero weight decay, using linear learning-rate warmup followed by linear decay.
We set the Jensen--Shannon mixing weight to $\beta = 0.5$, which yields the standard Jensen--Shannon divergence, and the grounding weight to $\gamma = 0.1$.
The trust region uses a maximum interpolation coefficient of $\lambda_{\max} = 1$ and a divergence budget of $\eps = 0.02$.
At each token, we approximate the maximal feasible interpolation coefficient $\lambda^{\star}$ in Eq.~\ref{eq:lambda} by running eight iterations of bisection over $[0,1]$.
The resulting interval has width at most $1/2^8=1/256$.

\subsection{Prompt Structure}
\label{app:prompt-template}

Table~\ref{tab:prompt_template} summarizes the prompt structure used by \methodname.
The student receives only information available at the current decision, including the user context, executed history, current interface state, and executable actions.
Trajectory-level intent is appended only to the frozen teacher during training and is absent from student inference.

\begin{table}[!htb]
\caption{Prompt structure used by \methodname. Italicized fields in braces are filled for each decision.}
\label{tab:prompt_template}
\centering
\small
\setlength{\tabcolsep}{7pt}
\renewcommand{\arraystretch}{1.16}
\begin{tabular}{@{}p{0.19\textwidth}p{0.75\textwidth}@{}}
\toprule
Prompt block & Template content \\
\midrule
\multicolumn{2}{@{}l}{\textit{Shared student--teacher prompt}} \\
\addlinespace[2pt]
System instruction
& Role-play a real shopper, infer the next action from the supplied context, and return a concise first-person rationale followed by one structured action. \\
User context
& \{\textit{user\_profile}\}, \{\textit{recent\_purchases}\}, and \{\textit{recent\_clicks}\}. The two history lists contain up to 20 titles available before the interaction, with consecutive duplicates collapsed. \\
Interaction history
& \{\textit{interaction\_history}\} contains prior observations, rationales, and executed actions in chronological order. We retain the 10 most recent actions for single-session evaluation and 20 across visits for cross-session evaluation. \\
Current environment
& \{\textit{current\_screenshot}\}, \{\textit{page\_description}\}, \{\textit{visible\_elements}\}, and \{\textit{page\_state}\}. \\
Executable actions
& \{\textit{available\_actions}\} is a JSON list of actions accepted in the current state, including valid targets and the permitted scrolling range. \\
\addlinespace[3pt]
\multicolumn{2}{@{}l}{\textit{Teacher-only training context}} \\
\addlinespace[2pt]
Trajectory-level intent
& \{\textit{trajectory\_intent}\} contains high-level behavioral intent extracted from the complete training trajectory. It excludes evidence indices and is never exposed to the student. \\
\addlinespace[3pt]
\multicolumn{2}{@{}l}{\textit{Shared response contract}} \\
\addlinespace[2pt]
Response format
& \{``rationale'': ``\ldots'', ``action'': \{\ldots\}\}. The action must match one executable candidate. \\
\bottomrule
\end{tabular}
\end{table}

\section{Additional Experiment Details}
\label{app:additional-analyses}

\subsection{Production A/B Validation Details}
\label{app:ab-validation}

Section~\ref{subsec:ab-validation} validates simulator-based policy comparisons offline against outcomes recorded in a production A/B test.
The production reference is computed from outcomes recorded over a fixed observation window in two production traffic buckets, one serving the control policy and the other the treatment policy.
We evaluate the same pair of policies in simulation using a fixed pool of 10,000 users and a frozen product catalog.
Each policy is assigned exactly 5,000 users, and every simulator receives the same user assignment and product catalog.
We compare five task-specific checkpoints, including \methodname, STA, Customer-R1, Shop-R1, and AlignUSER, with the Qwen3.5-2B base model and two general-purpose foundation models, GPT-6-Sol and GPT-6-Luna.\footnote{Official model documentation: \url{https://developers.openai.com/api/docs/models/all}.}

We compute PV-CTR within each policy arm as the total number of click actions divided by the total number of exposure page views.
This definition matches the production metric and counts repeated exposures to the same item.
For both the recorded production results and the offline simulation, we define the policy effect as
$\Delta\text{PV-CTR}=\text{PV-CTR}_{\mathrm{treatment}}-\text{PV-CTR}_{\mathrm{control}}$.
The effect ratio divides the simulated policy effect by the production policy effect, and the effect error is the absolute deviation of this ratio from 1.

Table~\ref{tab:abtest-ctr} reveals three findings about simulator calibration to the production policy effect.
First, \methodname provides the closest match, achieving an effect ratio of 0.713$\times$ and an effect error of 0.287, which is 50.2\% lower than that of the next-closest simulator.
In contrast, the other task-specific simulators substantially attenuate the policy response, with effect ratios ranging from 0.020$\times$ to 0.297$\times$.
The untuned Qwen3.5-2B model reaches only 0.004$\times$.
Second, the general-purpose GPT-6-Sol and GPT-6-Luna models overestimate the production response, reaching 1.576$\times$ and 2.154$\times$, respectively.
This result indicates that general-purpose model capacity, without task-specific behavioral calibration, does not directly translate into realistic sensitivity to ranking-policy changes.
Third, all evaluated simulators retain some gap from the production reference.
Even \methodname, the closest method in this comparison, underestimates the production effect by 28.7\%.

\subsection{Efficiency Analysis}
\label{app:efficiency}
In this section, we compare how the inference cost of each method evolves over an interaction.
For every Rollout step $k\in\{1,\ldots,31\}$ of single-session evaluation, we record input, output, and total token counts together with model-call latency for each trajectory that reaches step $k$.
If a method makes multiple calls at the same Rollout step, including planning calls or invalid-action retries, we first sum their token counts and latency within that trajectory-step.
At each step, we then report the mean and sample standard deviation across the contributing trajectories.
This step-wise protocol separates cost growth caused by an expanding interaction history from differences in when simulators terminate.
All methods use the same inference hardware and evaluation configuration so that the latency comparison reflects method-level computation rather than different execution settings.

\begin{figure}[H]
    \centering
    \includegraphics[width=\textwidth]{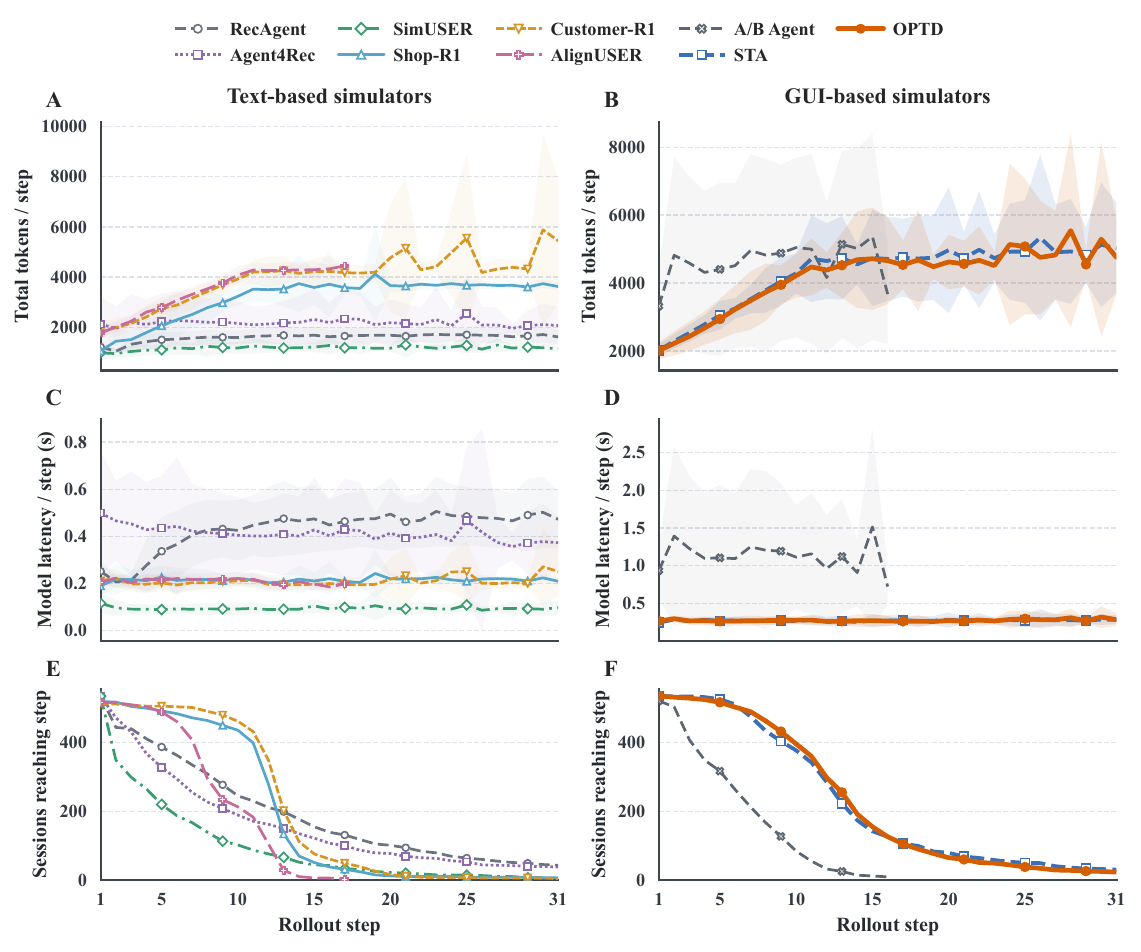}
    \caption{Step-wise inference cost on single-session Rollout. Lines show the mean across simulated trajectories that reach each step, and shaded bands span the mean $\pm 1$ sample standard deviation. Before aggregation, we sum all model calls assigned to the same trajectory-step, including method-internal calls and invalid-action retries. Panels A--B report total tokens, Panels C--D report model-call latency, and Panels E--F report the number of contributing trajectories. No absent step is imputed, so curves end when no trajectory reaches a later step.}
    \label{fig:rollout-efficiency}
\end{figure}

Figure~\ref{fig:rollout-efficiency} shows that the context-carrying GUI simulators exhibit the expected increase in token cost as interaction histories grow.
Among text-based simulators, AlignUSER has the highest mean token cost at earlier steps, while Customer-R1 becomes the most expensive among the methods that reach later steps.
Among GUI-based simulators, A/B Agent generally has the highest token cost and requires substantially more model calls per step than STA or \methodname.
\methodname generally uses fewer total tokens than STA, while their latency curves remain closely aligned.
Thus, the fidelity gains of \methodname do not require a systematic increase in per-step model latency.
Additionally, the token-cost shaded bands widen at later steps because the remaining trajectories have accumulated more heterogeneous interaction histories, resulting in greater variation in token cost.

\subsection{Qualitative Case Studies}
\label{app:case-studies}

We present representative examples to make the interaction structure of \textsc{UserArena} and the trajectory-level intent used by \methodname concrete.

\subsubsection{Representative User Trajectories}
\label{app:trajectory-cases}

\begin{figure}[H]
  \centering
  \includegraphics[width=\textwidth]{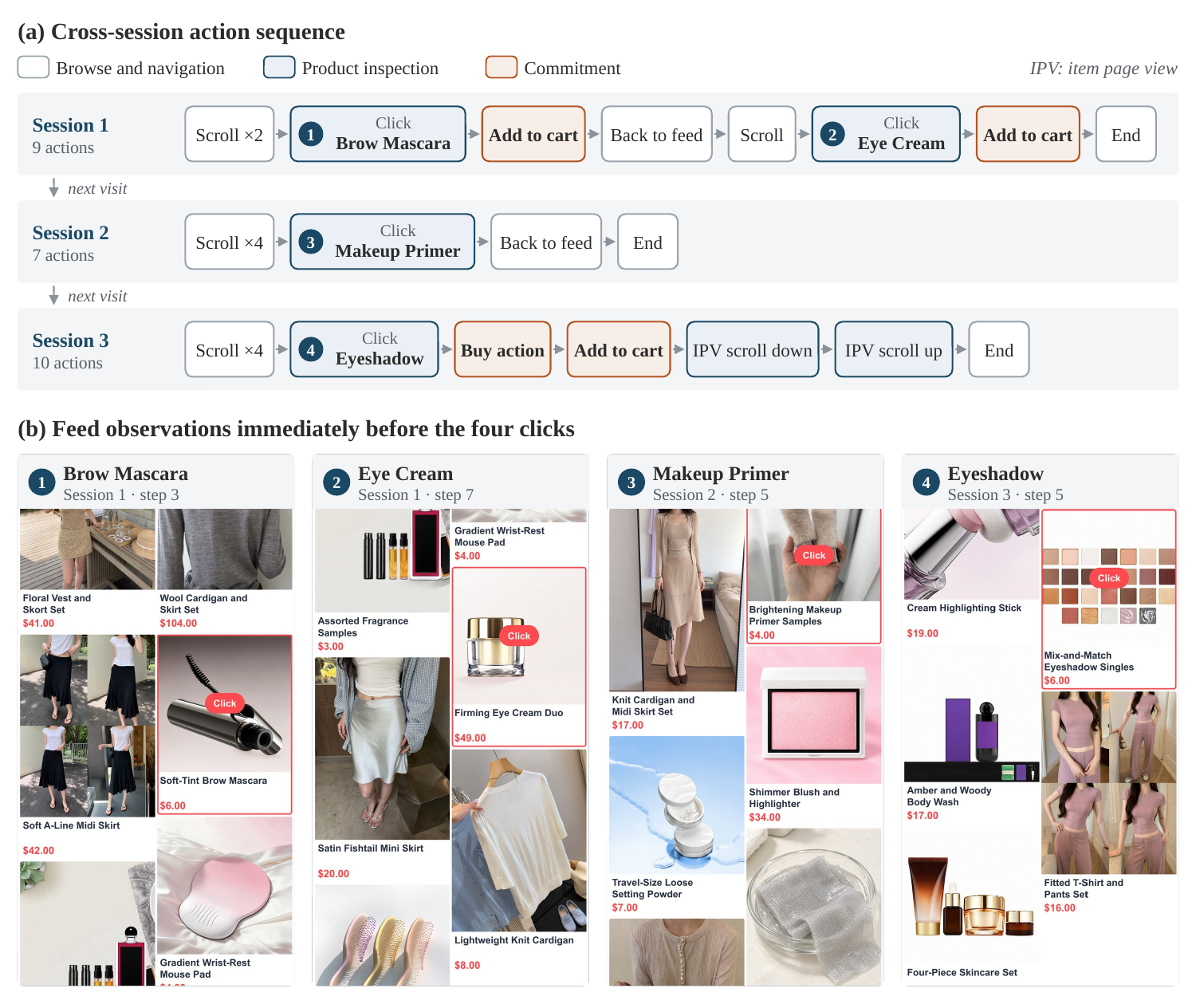}
  \caption[Ground-truth cross-session behavior over three consecutive visits by the same user.]{Ground-truth cross-session behavior over three consecutive visits by the same user. (a) Complete action sequence, with consecutive feed-scroll actions grouped as \emph{Scroll $\times N$} for readability. Colors distinguish browsing and navigation, product inspection, and commitment behavior. (b) Full feed observations immediately before four product clicks. Red outlines identify the selected products, and numbered markers link each observation to the corresponding click in panel (a).\protect\footnotemark}
  \label{fig:cross-session-case}
\end{figure}

Figure~\ref{fig:cross-session-case} shows one anonymized ground-truth trajectory from three consecutive visits by the same user.
Panel (a) preserves the order and count of all 26 actions, grouping only consecutive feed-scroll actions for readability.
Panel (b) presents the full feed observation immediately before each of four product clicks, and the numbered markers connect these observations to their positions in the action sequence.

\footnotetext{Product titles are translated for presentation. Item order, viewport position, clicked items, and action order are preserved.}

In Session 1, the user adds a brow mascara and an eye cream to the cart after alternating between feed navigation and product inspection.
Session 2 continues the broader interest in beauty products but ends after viewing a makeup primer.
Session 3 returns to eye makeup and records both \emph{Buy Action} and \emph{Add to Cart} for an eyeshadow product, followed by additional product-detail browsing.
Across the three visits, the user's broader interest in beauty products persists, while the product focus shifts from brow and eye care to face makeup and then back to eye makeup.
The example therefore illustrates how \textsc{UserArena} preserves both continuity and modest shifts in user interests, together with changing levels of engagement, across chronologically ordered sessions.

\subsubsection{Trajectory-Level Intent Example}
\label{app:intent-cases}

\begin{figure}[H]
  \centering
  \includegraphics[width=\textwidth]{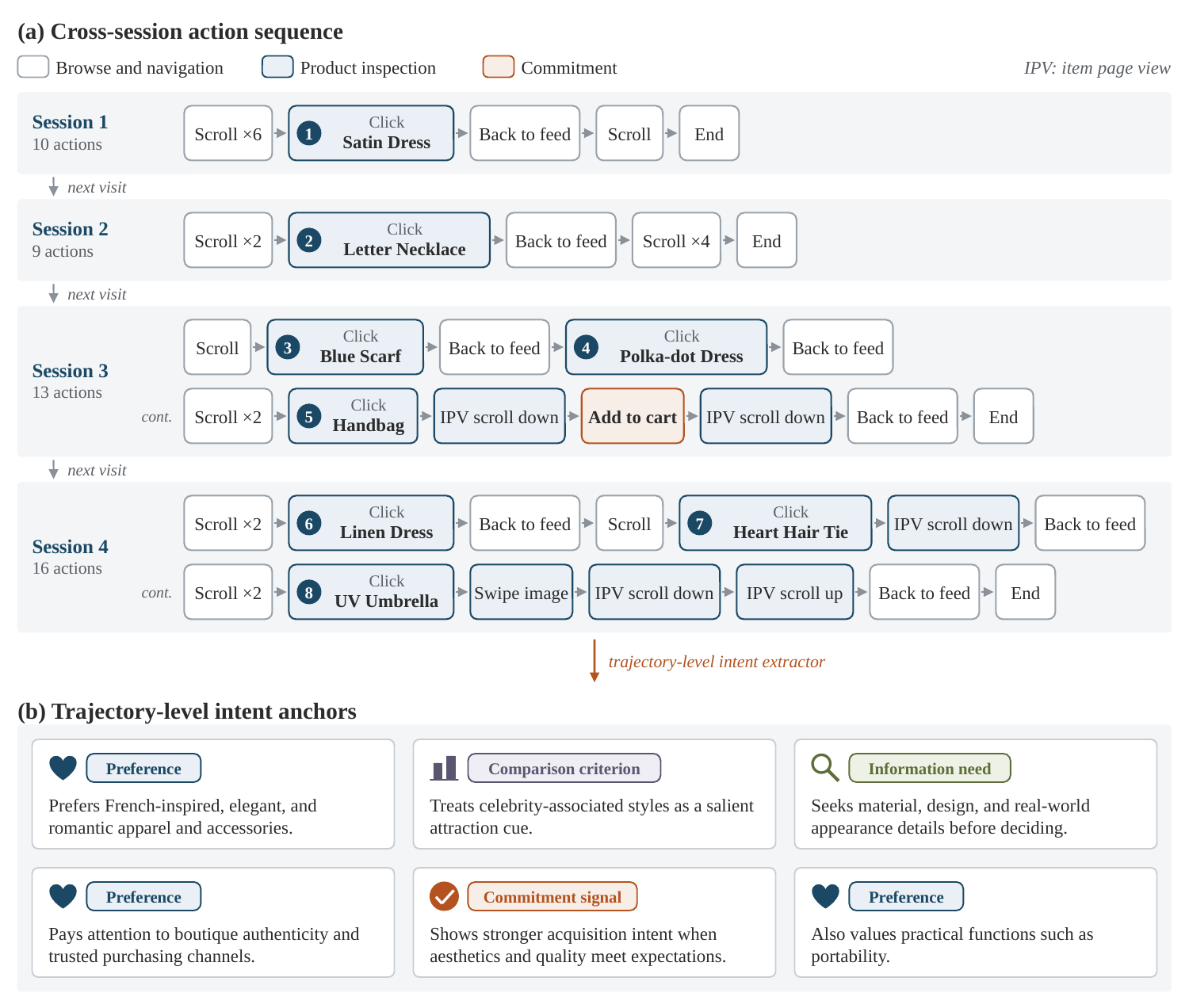}
  \caption[One cross-session trajectory and its structured trajectory-level intent.]{One cross-session trajectory and its structured trajectory-level intent.
(a) Complete behavior across four consecutive visits by the same user.
Colors distinguish browsing and navigation, product inspection, and commitment behavior.
Consecutive feed-scroll actions are grouped as \emph{Scroll $\times N$}, and the two longer visits wrap onto a continuation line for readability.
(b) The single structured intent record extracted from the complete trajectory.
Each card is a typed anchor within the same intent.}
  \label{fig:intent-case}
\end{figure}

Figure~\ref{fig:intent-case} pairs a real-user training trajectory with the privileged trajectory-level intent extracted from it and supplied to the frozen teacher.
Panel~(a) preserves all 48 recorded actions, grouping only consecutive feed-scroll actions and wrapping the two longer sessions onto continuation lines.
Panel~(b) shows the resulting intent as it is represented during training: one structured record intent containing six complementary typed anchors.

Across the four visits, the user repeatedly inspects premium apparel and accessories, while varying the product category from dresses and jewelry to a scarf, handbag, hair accessory, and umbrella.
The extracted preferences capture the stable interest in elegant design and branded goods.
The comparison criterion abstracts the recurring attention to celebrity-associated styles, while the commitment signal reflects stronger acquisition intent when aesthetic and quality expectations align.
None of these anchors identifies a specific product, page, price, or next action.
The same structured record conditions the teacher at every state of this trajectory, while the student receives neither $z_u$ nor future observations.
The teacher must therefore guide the student's own response prefix at the current state, preserving the on-policy supervision used to reduce the mismatch between teacher-forced training and autoregressive inference.

\subsection{Turing-Style Human Evaluation Details}
\label{app:turing-interface}

\begin{figure}[H]
  \centering
  \includegraphics[width=0.81\textwidth]{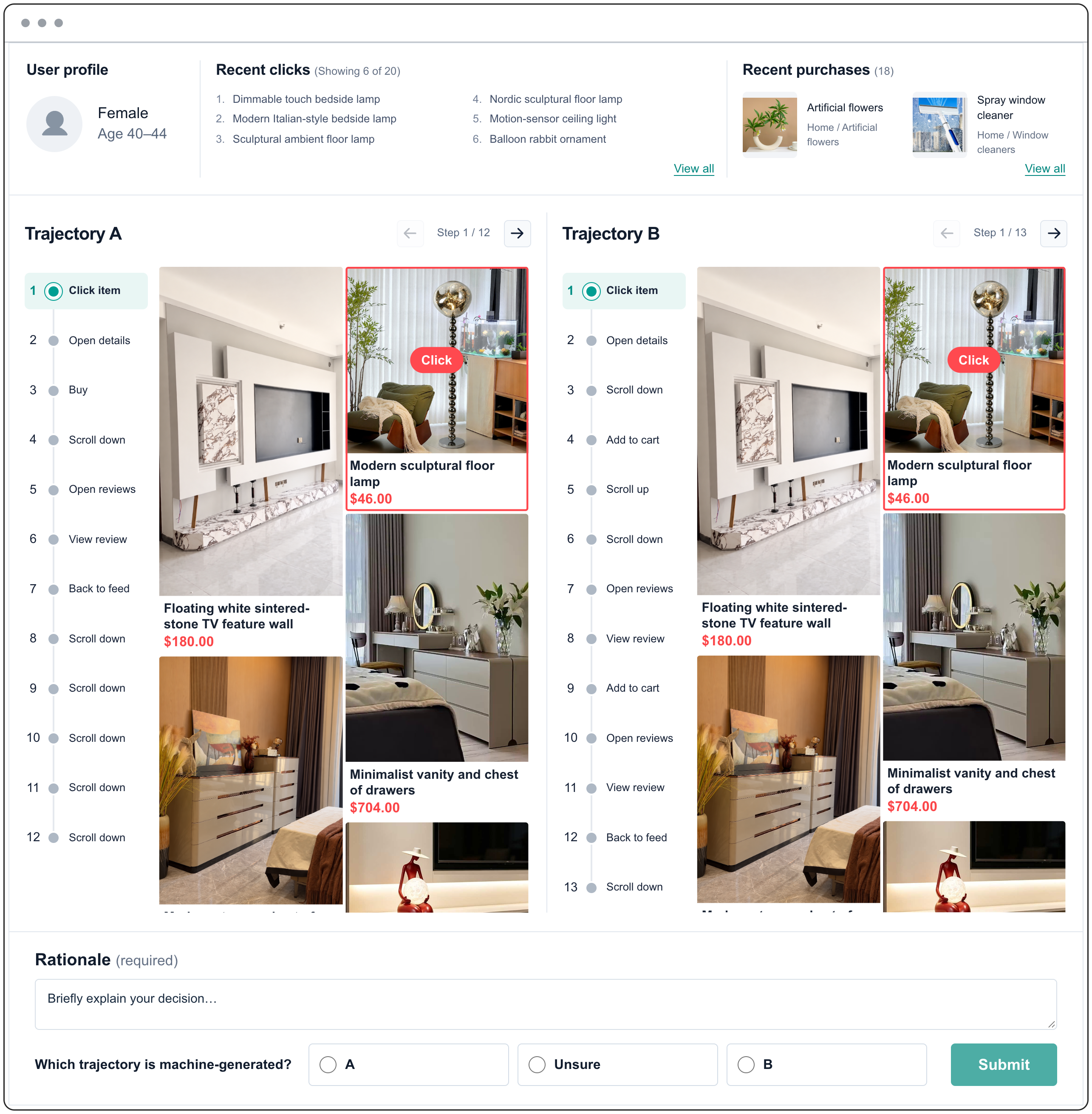}
  \caption{Human evaluation interface. Annotators compare two anonymized e-commerce interaction trajectories conditioned on the same user context. The interface presents recent interaction history, action timelines, and step-level interface observations. Annotators select the trajectory they believe is machine-generated or choose \emph{Unsure}, and provide a supporting rationale.}
  \label{fig:turing-interface}
\end{figure}

The source-blinded evaluation in Section~\ref{subsec:in-depth} compares \methodname with Customer-R1, STA, and A/B Agent.
For each user context, we pair a real trajectory with one generated by a simulator, anonymize the sources as A and B, and randomize their display order.
Three annotators independently inspect the action sequences and corresponding interface observations, then choose \emph{A is machine-generated}, \emph{B is machine-generated}, or \emph{Unsure}.
Each simulator is evaluated on 200 trajectory pairs sampled from the held-out test set.
\textbf{Annotator characteristics.}
All annotators are adults (18+) familiar with e-commerce shopping interfaces and have experience with recommender systems or user-behavior annotation.
All annotators provided informed consent and were compensated at prevailing local market rates.
Recruitment considered non-sensitive demographic characteristics, educational background, and relevant domain expertise.
No sensitive personal attributes were collected or used for annotator selection or analysis.
Figure~\ref{fig:turing-interface} shows the shared user context, both action timelines and their step-level interface observations, the three response options, and the required rationale field.

Figure~\ref{fig:ablation-turing}(b) shows that \methodname has the lowest Human win rate (38.7\%) and the highest Model win rate (26.3\%) among the four simulators, while Customer-R1 has the highest Unsure rate (47.0\%).
Across trajectory pairs, Fleiss' $\kappa$ is 0.528~\citep{fleiss1971measuring}.
This statistic measures consistency among annotators rather than attribution correctness or human-likeness.
Together, the agreement score and source-attribution rates indicate that annotators often reach similar judgments while still finding \methodname trajectories difficult to distinguish from recorded behavior.
This pattern suggests that \textsc{UserArena}'s densely observed interactions provide useful supervision for learning human-like trajectories.

\end{document}